\documentstyle[twoside,epsf]{article}
\catcode`\@=11
\long\def\@makefntext#1{
\protect\noindent \hbox to 3.2pt {\hskip-.9pt  
$^{{\eightrm\@thefnmark}}$\hfil}#1\hfill}		

\def\thefootnote{\fnsymbol{footnote}}
\def\@makefnmark{\hbox to 0pt{$^{\@thefnmark}$\hss}}	
	
\def\ps@myheadings{\let\@mkboth\@gobbletwo
\def\@oddhead{\hbox{}
\rightmark\hfil\eightrm\thepage}   
\def\@oddfoot{}\def\@evenhead{\eightrm\thepage\hfil
\leftmark\hbox{}}\def\@evenfoot{}
\def\sectionmark##1{}\def\subsectionmark##1{}}

\oddsidemargin=\evensidemargin
\renewcommand{\thefootnote}{\fnsymbol{footnote}}

\newcounter{sectionc}\newcounter{subsectionc}\newcounter{subsubsectionc}
\renewcommand{\section}[1] {\vspace{12pt}\addtocounter{sectionc}{1} 
\setcounter{subsectionc}{0}\setcounter{subsubsectionc}{0}\noindent 
	{\tenbf\thesectionc. #1}\par\vspace{5pt}}
\renewcommand{\subsection}[1] {\vspace{12pt}\addtocounter{subsectionc}{1} 
	\setcounter{subsubsectionc}{0}\noindent 
	{\bf\thesectionc.\thesubsectionc. {\kern1pt \bfit #1}}\par\vspace{5pt}}
\renewcommand{\subsubsection}[1] {\vspace{12pt}\addtocounter{subsubsectionc}{1}
	\noindent{\tenrm\thesectionc.\thesubsectionc.\thesubsubsectionc.
	{\kern1pt \tenit #1}}\par\vspace{5pt}}
\newcommand{\nonumsection}[1] {\vspace{12pt}\noindent{\tenbf #1}
	\par\vspace{5pt}}

\newcounter{appendixc}
\newcounter{subappendixc}[appendixc]
\newcounter{subsubappendixc}[subappendixc]
\renewcommand{\thesubappendixc}{\Alph{appendixc}.\arabic{subappendixc}}
\renewcommand{\thesubsubappendixc}
	{\Alph{appendixc}.\arabic{subappendixc}.\arabic{subsubappendixc}}

\renewcommand{\appendix}[1] {\vspace{12pt}
        \refstepcounter{appendixc}
        \setcounter{figure}{0}
        \setcounter{table}{0}
        \setcounter{lemma}{0}
        \setcounter{theorem}{0}
        \setcounter{corollary}{0}
        \setcounter{definition}{0}
        \setcounter{equation}{0}
        \renewcommand{\thefigure}{\Alph{appendixc}.\arabic{figure}}
        \renewcommand{\thetable}{\Alph{appendixc}.\arabic{table}}
        \renewcommand{\theappendixc}{\Alph{appendixc}}
        \renewcommand{\thelemma}{\Alph{appendixc}.\arabic{lemma}}
        \renewcommand{\thetheorem}{\Alph{appendixc}.\arabic{theorem}}
        \renewcommand{\thedefinition}{\Alph{appendixc}.\arabic{definition}}
        \renewcommand{\thecorollary}{\Alph{appendixc}.\arabic{corollary}}
        \renewcommand{\theequation}{\Alph{appendixc}.\arabic{equation}}
        \noindent{\tenbf Appendix \theappendixc #1}\par\vspace{5pt}}
\newcommand{\subappendix}[1] {\vspace{12pt}
        \refstepcounter{subappendixc}
        \noindent{\bf Appendix \thesubappendixc. {\kern1pt \bfit #1}}
	\par\vspace{5pt}}
\newcommand{\subsubappendix}[1] {\vspace{12pt}
        \refstepcounter{subsubappendixc}
        \noindent{\rm Appendix \thesubsubappendixc. {\kern1pt \tenit #1}}
	\par\vspace{5pt}}

\newcommand{\textlineskip}{\baselineskip=13pt}
\newcommand{\smalllineskip}{\baselineskip=10pt}

\def\eightcirc{
\begin{picture}(0,0)
\put(4.4,1.8){\circle{6.5}}
\end{picture}}
\def\eightcopyright{\eightcirc\kern2.7pt\hbox{\eightrm c}} 

\newcommand{\copyrightheading}[1]
	{\vspace*{-2.5cm}\smalllineskip{\flushleft
	{\footnotesize International Journal of Modern Physics A, #1}\\
	{\footnotesize $\eightcopyright$\, World Scientific Publishing
	 Company}\\
	 }}

\newcommand{\publisher}[2]{{\begin{center}\footnotesize\smalllineskip 
	Received #1\\
	Revised #2
	\end{center}
	}}

\def\abstracts#1#2#3{{
	\centering{\begin{minipage}{4.5in}\baselineskip=10pt\footnotesize
	\parindent=0pt #1\par 
	\parindent=15pt #2\par
	\parindent=15pt #3
	\end{minipage}}\par}} 

\renewenvironment{thebibliography}[1]
	{\frenchspacing
	 \ninerm\baselineskip=11pt
	 \begin{list}{\arabic{enumi}.}
	{\usecounter{enumi}\setlength{\parsep}{0pt}
	 \setlength{\leftmargin 12.7pt}{\rightmargin 0pt} 
	 \setlength{\itemsep}{0pt} \settowidth
	{\labelwidth}{#1.}\sloppy}}{\end{list}}

\newcounter{itemlistc}
\newcounter{romanlistc}
\newcounter{alphlistc}
\newcounter{arabiclistc}

\newcommand{\fcaption}[1]{
        \refstepcounter{figure}
        \setbox\@tempboxa = \hbox{\footnotesize Fig.~\thefigure. #1}
        \ifdim \wd\@tempboxa > 5in
           {\begin{center}
        \parbox{5in}{\footnotesize\smalllineskip Fig.~\thefigure. #1}
            \end{center}}
        \else
             {\begin{center}
             {\footnotesize Fig.~\thefigure. #1}
              \end{center}}
        \fi}

\newcommand{\tcaption}[1]{
        \refstepcounter{table}
        \setbox\@tempboxa = \hbox{\footnotesize Table~\thetable. #1}
        \ifdim \wd\@tempboxa > 5in
           {\begin{center}
        \parbox{5in}{\footnotesize\smalllineskip Table~\thetable. #1}
            \end{center}}
        \else
             {\begin{center}
             {\footnotesize Table~\thetable. #1}
              \end{center}}
        \fi}

\def\@citex[#1]#2{\if@filesw\immediate\write\@auxout
	{\string\citation{#2}}\fi
\def\@citea{}\@cite{\@for\@citeb:=#2\do
	{\@citea\def\@citea{,}\@ifundefined
	{b@\@citeb}{{\bf ?}\@warning
	{Citation `\@citeb' on page \thepage \space undefined}}
	{\csname b@\@citeb\endcsname}}}{#1}}

\newif\if@cghi
\def\cite{\@cghitrue\@ifnextchar [{\@tempswatrue
	\@citex}{\@tempswafalse\@citex[]}}
\def\citelow{\@cghifalse\@ifnextchar [{\@tempswatrue
	\@citex}{\@tempswafalse\@citex[]}}
\def\@cite#1#2{{$\null^{#1}$\if@tempswa\typeout
	{IJCGA warning: optional citation argument 
	ignored: `#2'} \fi}}

\def\pmb#1{\setbox0=\hbox{#1}
	\kern-.025em\copy0\kern-\wd0
	\kern.05em\copy0\kern-\wd0
	\kern-.025em\raise.0433em\box0}

\def\fnm#1{$^{\mbox{\scriptsize #1}}$}
\def\fnt#1#2{\footnotetext{\kern-.3em
	{$^{\mbox{\scriptsize #1}}$}{#2}}}

\def\fpage#1{\begingroup
\voffset=.3in
\thispagestyle{empty}\begin{table}[b]\centerline{\footnotesize #1}
	\end{table}\endgroup}

\def\runninghead#1#2{\pagestyle{myheadings}
\markboth{{\protect\footnotesize\it{\quad #1}}\hfill}
{\hfill{\protect\footnotesize\it{#2\quad}}}}
\font\tenrm=cmr10
\font\tenit=cmti10 
\font\tenbf=cmbx10
\font\bfit=cmbxti10 at 10pt
\font\ninerm=cmr9

\font\eightrm=cmr8

\def\qed{\hbox{${\vcenter{\vbox{			
   \hrule height 0.4pt\hbox{\vrule width 0.4pt height 6pt
   \kern5pt\vrule width 0.4pt}\hrule height 0.4pt}}}$}}

\renewcommand{\thefootnote}{\fnsymbol{footnote}}	

\newlength{\mywidth}\mywidth=2.75truein 
\def\fref#1{рис.\ref{#1}}

\renewenvironment{figure}{\refstepcounter{figure}
\baselineskip=0.4\normalbaselineskip\footnotesize}
{\baselineskip=\normalbaselineskip}
\newenvironment{figtr}{}{}
\def\fignum{{\bf Рис.\arabic{chapter}.\arabic{figure}.\quad}}

\def\nn{\nonumber}

\def\s{\sigma}
\def\half{{\textstyle{{1}\over{2}}}}
\def\quart{{\textstyle{{1}\over{4}}}}

\def\Z{{\bf Z}}

\def\la{\bigl}
\def\ra{\bigr}

\def\intt{\int_{0}^{2\pi}d\s}

\newcommand{\q}{Q_{\mu}(\sigma)}

\newcommand{\p}{P_{\mu}}

\newcommand{\sqp}{\sqrt{P^{2}}}
\newcommand{\astar}{a^{*}}

\newcommand{\e}{{\bf e}_{i}}
\newcommand{\sbo}{{\bf S}}

\newcommand{\nm}{n_{\mu}^{-}}

\newcommand{\ints}{\int\limits_{0}^{2\pi}d\sigma}

\newcommand{\kr}{\Gamma _{\mu}^{ij}=N_{\nu}^{i}\partial N_{\nu}^{j}/
   \partial \p}  
\newcommand{\sdef}{\sbo=-\frac{1}{4}\ints\int\limits_{0}^{\sigma} d\sigma '\;
    {\bf a}(\sigma)\times {\bf a}(\sigma ')}

\newcommand{\hi}[1]{\chi_{#1}}

\newcommand{\es}[1]{{\bf e}_{#1}}

\newcommand{\pz}{\p,Z_{\mu}}
\newcommand{\akak}{a_{k},a_{k}^{*}}

\newcommand{\df}{\partial}

\def\frac#1#2{{\textstyle{{#1}\over{#2}}}}

\def\fignum{{\bf Fig.\arabic{figure}.\quad}}
\def\fref#1{fig.\ref{#1}}

\begin{document}

\runninghead{String theory in Lorentz-invariant light cone gauge} {String theory in Lorentz-invariant light cone gauge} 

\normalsize\textlineskip
\thispagestyle{empty}
\setcounter{page}{1}

\copyrightheading{}			

\vspace*{0.88truein}

\fpage{1}
\centerline{\bf STRING THEORY IN LORENTZ-INVARIANT LIGHT CONE GAUGE}
\vspace*{0.37truein}
\centerline{\footnotesize IGOR NIKITIN\footnote{E-mail: Igor.Nikitin@gmd.de}}
\vspace*{0.015truein}
\centerline{\footnotesize\it Natinal Research Center for 
Information Technology,}
\centerline{\footnotesize\it  GMD/VMSD, 53754 St.Augustin, Germany}
\vspace*{0.225truein}
\publisher{}{}

\vspace*{0.21truein}
\abstracts{Quantization of 4-dimensional Nambu-Goto theory of open string
in light cone gauge, related in Lorentz-invariant way with the
world sheet, is performed. Obtained quantum theory has no
anomalies in Lorentz group. Determined spin-mass spectra
of the theory have Regge-like behavior and do not contain
the tachyon. Vertex operators of interaction theory,
acting in the physical subspace, are constructed.
}{}{}


\vspace*{1pt}\textlineskip	
\vspace*{1cm}
\noindent
It is  well known, that quantization of string theory 
creates anomalies, destroying classical symmetries of the system. 
Exception is a theory at critical value of space-time dimension $d=26$,
where the anomalies are absent. To construct 4-dimensional quantum string 
theory, required in physical applications, one usually introduces
additional degrees of freedom (such as fermion fields, propagating
along the world sheet), whose contribution cancels the anomaly at
less number of dimensions. This approach can be combined with another idea:
some of dimensions can be considered as coordinates on a compact
manifold of physically negligible size. 

There are also other methods of string quantization, 
which do not introduce extra dimensions or extra degrees of freedom,
but use non-standard canonical bases in the phase space of Hamiltonian 
mechanics as a starting point for quantization. Such methods
were applied in works \cite{slstring,ax} for quantization of special
subsets in the phase space, representing particular types of the world 
sheets. Distinctive features of these 4-dimensional theories
are absence of anomalies in Lorentz group and non-fixed intercept
in spin-mass spectrum, which gives a possibility to remove tachyon
from the theory. Due to these features, the given approach can be used
for a construction of relativistic models of hadrons \cite{Berd}.


In this work we will try to
extend the methods \cite{slstring,ax} for the world sheets
of general form. For this purpose we use a definite 
modification of light cone gauge.

Light cone gauge relates a parametrization of the world sheet
with some light-like vector (gauge axis), see \fref{f0}. 
In standard approach this vector is non-dynamical, e.g.
$n_{\mu}=(1,1,0,0...)$. Because the Lorentz-transformations 
change the position of the world sheet respective to this fixed
axis, they are followed by reparametrizations of the world sheet.
On quantum level the reparametrization group has anomaly,
which appears also in Lorentz group and violates
Lorentz covariance of the theory. This is a main problem
of string theory in standard light cone gauge.

The simple idea how to avoid this problem  
is to connect the gauge axis with some {\it dynamical} vector
in string theory. In this case
the Lorentz-transformations move the gauge axis together with the
world sheet, and the parametrization on the world sheet is not changed.
Lorentz group in this approach should be free of anomalies.

\textheight=7.8truein
\setcounter{footnote}{0}
\renewcommand{\thefootnote}{\alph{footnote}}

\parbox[t]{5cm}{
\begin{figure}\label{f0}
\begin{center}
~\epsfysize=3.6cm\epsfxsize=5cm\epsffile{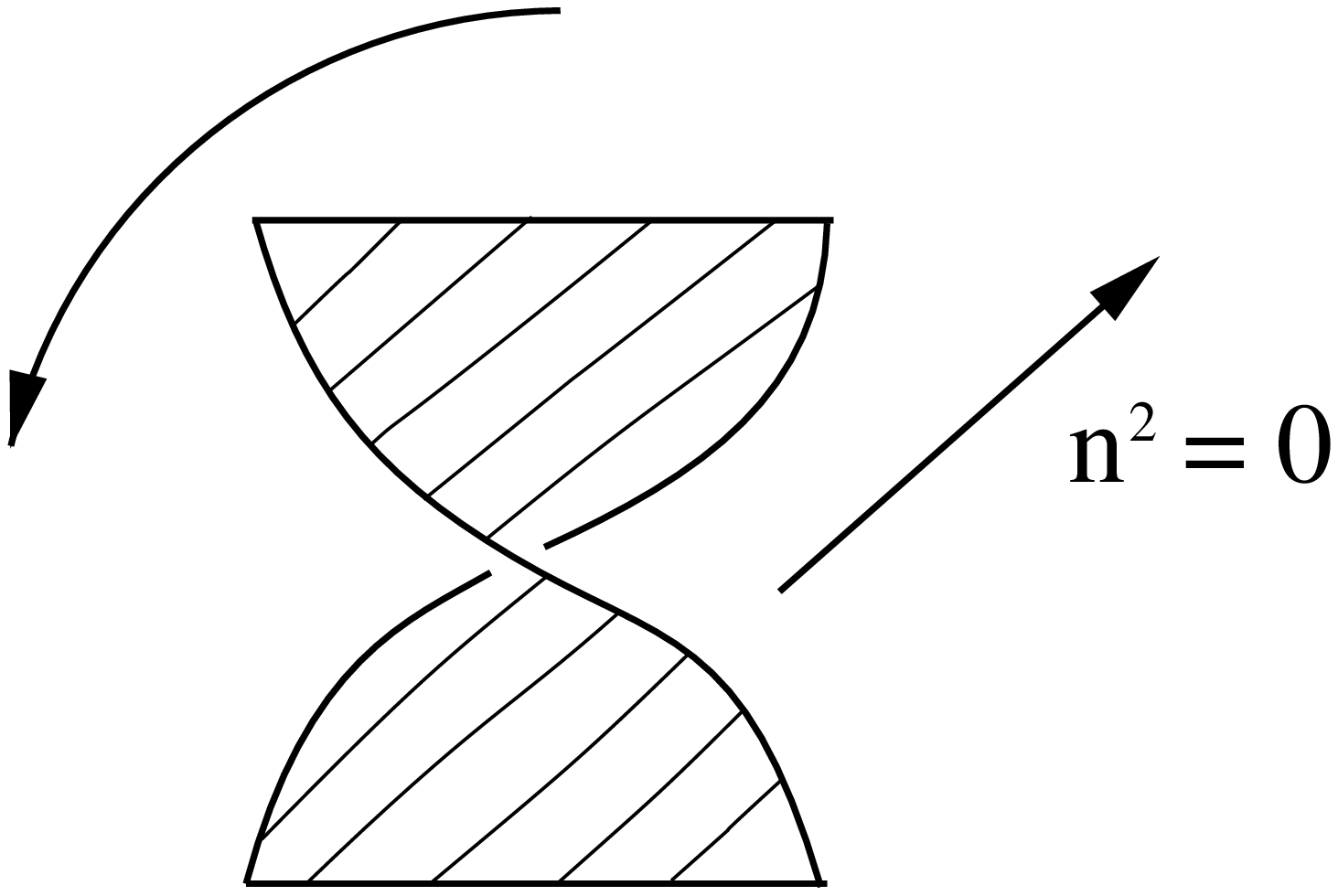}

\vspace{5mm}
\fignum Light cone gauge.
\end{center}
\end{figure}
}\quad\quad\parbox[t]{6cm}{
\begin{figure}\label{f1}
\begin{center}
~\epsfysize=4.5cm\epsfxsize=6cm\epsffile{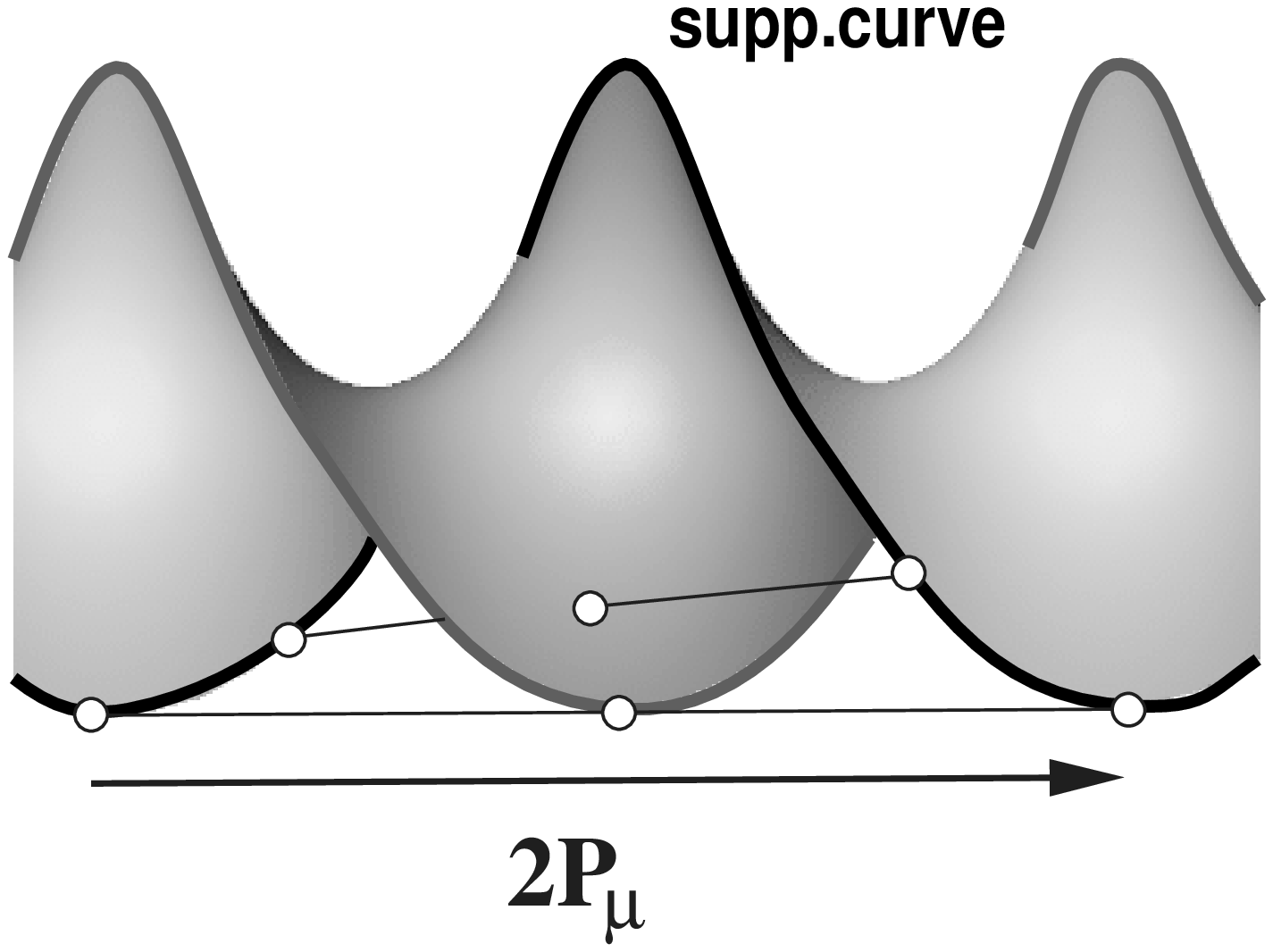}

\parbox{6cm}{\fignum World sheet of open string is constructed as
a locus of middles of segments,
connecting all possible pairs of points on the supporting curve.
}
\end{center}
\end{figure}
}

\vspace{5mm}
\noindent Implementation of this idea includes the following ingredients.

\paragraph*{Geometrical description of the world sheet \cite{zone}.}
Let's introduce a function, related with
string's coordinates and momenta
by expressions

\vspace{-5mm}
\begin{eqnarray}
&&Q_{\mu}(\s)=x_{\mu}(\s)+\int_{0}^{\s}d\tilde\s \;
p_{\mu}(\tilde\s),\label{Qini0}\\
&&x_{\mu}(\s)=(Q_{\mu}(\s)+Q_{\mu}(-\s))/2,\quad
p_{\mu}(\s)=(Q'_{\mu}(\s)+Q'_{\mu}(-\s))/2\label{Qini}
\end{eqnarray}
($x,p$ are {\it even} functions of $\s$).
In terms of oscillator variables, commonly used in string theory:
\begin{eqnarray}
&&Q_{\mu}(\s)=X_{\mu}+\frac{P_{\mu}}{\pi}\sigma+
{\textstyle{{1}\over{\sqrt{\pi}}}}
\sum\limits_{n\neq0}{\textstyle{{a_{\mu}^{n}}\over{in}}}e^{in\sigma}.\label{Qosc}
\end{eqnarray}
The curve, defined by the function $Q_{\mu}(\s)$ (further called
{\it supporting curve}) has the following properties:

\vspace{3mm}

\noindent 1.~the curve is light-like: $Q'^{2}(\s)=0$, 
this property is equivalent to Virasoro constraints on oscillator variables;

\vspace{1mm}
\noindent 2.~the curve is periodical: $Q(\s+2\pi)-Q(\s)=2P$;

\vspace{1mm}
\noindent 3.~the curve coincides with the world line of one string end:
$x(0,\tau)=Q(\tau)$; the world line of another end
is the same curve, shifted onto the semi-period:
$x(\pi,\tau)=Q(\pi+\tau)-P$;

\vspace{1mm}
\noindent 4.~the whole world sheet is reconstructed by this curve
as follows: $x(\s,\tau)=(Q(\s_{1})+Q(\s_{2}))/2$, $\s_{1,2}=\tau\pm\s$,
see \fref{f1};

\vspace{1mm}
\noindent 5.~Poisson brackets for the function $Q_{\mu}(\s)$, and symplectic form,
correspondent to these brackets (see Appendix~1):
\begin{eqnarray}
&&\{Q_{\mu}(\s),Q_{\nu}(\tilde\s)\}=-2g_{\mu\nu}\vartheta(\s-\tilde\s),
\label{orig}\\
&&\Omega=\half\; dP_{\mu}\wedge dQ_{\mu}(0)+
\quart\intt\;\delta Q'_{\mu}(\s)\wedge\delta Q_{\mu}(\s).\nn
\end{eqnarray}
Here $\vartheta(\s)=[\s /2\pi]+\half$, $[x]$ is integer part of $x$,
a derivative $\vartheta(\s)'=\Delta(\s)$ is periodical delta-function.

\vspace{2mm}
These properties can be easily proven from definition of $Q_{\mu}(\s)$
and known mechanics in oscillator variables, see Appendix~2.

\paragraph*{Mechanics in center-of-mass frame \cite{slstring}.}
Let's introduce orthonormal tetrad of vectors, 
dependent on total momentum: $N^{\alpha}_{\mu}(P),\
N^{\alpha}_{\mu}N^{\beta}_{\mu}=g^{\alpha\beta}$, with
$N^{0}_{\mu}=P_{\mu}/\sqrt{P^{2}}$. Let's decompose the supporting curve 
by this tetrad: $Q_{\mu}(\s)=N^{\alpha}_{\mu}Q^{\alpha}(\s).$

\paragraph*{Lorentz-invariant light cone gauge \cite{ax}.}~
Virasoro constraints generate re\-pa\-ra\-me\-tri\-za\-ti\-ons 
of supporting curve (see Appendix~2). Gauges to Virasoro constraints
select particular parametrization on this curve. 

Let's use a parametrization:\quad
$Q^{\alpha}(\sigma) = Q^{\alpha}(0)+\int_{0}^{\sigma}
     d\sigma ' a^{\alpha}(\sigma '),$
\begin{eqnarray}
a^{\alpha}(\sigma) &=&  \left(
 \frac{\pi}{2\sqp}\left(\frac{P^{2}}
  {\pi^{2}}+|a(\sigma)|^{2}\right), 
\ \frac{a(\sigma)+\astar(\sigma)}{2} \; {\bf e}_{1} +
    \frac{a(\sigma)-\astar(\sigma)}{2} \; i{\bf e}_{2} \right.\label{conf1}\\
&&+\left.\frac{\pi}{2\sqp}\left(\frac{P^{2}}{\pi^{2}}-|a(\sigma)|^{2}
\right){\bf e}_{3} \right),\nn
\end{eqnarray}
where $a(\sigma)=\sqrt{\frac{2}{\pi}}\sum\limits_{n\neq0} a_{n}e^{-in\sigma}$
and ${\bf e}_{k}$ is an orthonormal basis in CMF.
Here one easily recognizes the light cone gauge
$Q^{+}(\sigma)\equiv \nm\q=Q^{+}(0)+\frac{1}{\pi}
     \sqrt{\frac {P^{2}}{2}}\;\sigma$ with the gauge axis
$\nm=\frac{1}{\sqrt{2}}(N_{\mu}^{0}-N_{\mu}^{i}e_{3}^{i})$.
The difference from standard approach is that $\nm$ is now {\it dynamical}
vector, because we interpret ${\bf e}_{k}$ as dynamical variables.

Substituting this parametrization into general symplectic form (\ref{orig}),
we obtain

\vspace{-5mm}\begin{eqnarray}
&&\Omega=d\p\wedge dZ_{\mu}\;+
 \sum_{k \neq 0}\frac{1}{ik}\; d\astar_{k}
 \wedge da_{k}
\; + \; \half \; d\e \wedge d(\sbo\times\e), \label{Olcg}
\end{eqnarray}

\vspace{-3mm}\noindent where 
$Z_{\mu}= {\textstyle{{1}\over{2\sqrt{P^{2}}}}}\ints a^{0}(\sigma)\biggl(
   Q_{\mu}(\sigma)-\biggl({\textstyle{{\sigma}\over{\pi}}}
-1\biggr)P_{\mu}\biggr)+
   \half \epsilon^{ijk}\Gamma _{\mu}^{ij}S^{k}$
is mean coordinate, conjugated to $P_{\mu}$; $\kr$ -- Christoffel symbols
and $\sdef$ is an orbital moment of the string in CMF (further called spin).
The  last   expression can  be  written  in  terms of
oscillators $a_{k}$  and can be  interpreted as {\it constraints} : 
\begin{eqnarray}
&&\chi_{3}=S_{3}-A_{3}=0,\quad\chi^{+}= S^{+}-A^{+}=0,\quad 
\chi^{-}=S^{-}-A^{-}=0,\quad
\mbox{where}\nn\\
&&S_{i}=S^{k}e_{i}^{k}\ 
\mbox{is a projection of ${\bf S}$ onto ${\bf e}_{i}$};\
 S^{\pm}=S_{1}\pm iS_{2};\ \chi^{\pm}=\chi_{1}\pm i\chi_{2};\nn\\
&&A_{3}=\sum\limits_{n\neq0} 
\frac{1}{n}a_{n}^{*}a_{n},\quad A^{-}=\sqrt{\frac{2\pi}{P^{2}}}
     \sum_{k,n,k+n\neq0} 
\frac{1}{k}a_{k}a_{n}a_{k+n}^{*},\quad A^{+}=\ \mbox{c.c.}\label{Apol}
\end{eqnarray}

Another constraint is given by the
requirement  of  $2P$-periodicity of the  curve:
\begin{eqnarray}
&& 2\p=\int_{0}^{2\pi}d\s\; a_{\mu}(\sigma)\ \Leftrightarrow\ \chi_{0}=
\frac{P^{2}}{2\pi}-L_{0}=0,\quad L_{0}=\sum\limits_{n\neq0} a_{n}^{*}a_{n}.\nn
\end{eqnarray}

The obtained symplectic form corresponds to the following Poisson brackets:
\begin{eqnarray}
&&\{Z_{\mu},P_{\nu}\}=g_{\mu\nu},  \ \{a_{k},a_{n}^{*}\}=ik\delta_{kn},
\ k,n\in\Z/\{0\},\label{Pb1}\\
&&\{S^{i},S^{j}\}=-\epsilon^{ijk}S^{k},  \
    \{S^{i},e^{j}_{n}\}=-\epsilon^{ijk}e^{k}_{n}. \nn
    \end{eqnarray}
Algebra of constraints belongs the  1st    class: 
$\{\hi{0},\hi{i}\}=0\ ,\ 
\{\hi{i},\hi{j}\}=\epsilon_{ijk}\hi{k}$.

\vspace{2mm}
 Thus, the string  is equivalent to a mechanical system:

\vspace{1mm}
\begin{center}
$\pz$ +  infinite set  of  oscillators $\akak$ +  the top $\e,{\bf S}$,
\end{center}

\vspace{1mm}\noindent
restricted by 4 constraints of the 1st class,
which include mass shell condition and requirements
of the form ``spin of the top is equal to the spin of the string''.  
Constraints generate reparametrizations of the supporting curve:
$\chi_{0}$ generates shift of argument $Q(\s)\to Q(\s+\tau)$
(evolution of the string, see Appendix~2); $\chi_{i}$ generate the rotations of basis $\e$   
with respect to non-moving supporting curve. 

In more details: $\chi_{3}$
rotates $\es{1}$ and $\es{2}$ about $\es{3}$ and simultaneously
rotates coefficients of supporting curve decomposition 
in opposite direction, so that supporting curve is not changed:
$\{\chi_{3},Q(\s)\}=0$. Constraints $\chi_{1,2}$ generate the  rotations 
of basis, changing the direction of the gauge axis $n_{\mu}^{-}$,
and they are followed by reparametrization of the supporting curve.
In the  oscillator variables  this  reparametrization  looks  
like a complicated nonlinear transformation.

Generators of Lorentz group are defined by expression \cite{slstring}
\begin{eqnarray}
&&M_{\mu\nu}=\int_{0}^{\pi}d\s(x_{\mu}p_{\nu}-x_{\nu}p_{\mu})
=X_{\mu}P_{\nu}-X_{\nu}P_{\mu}+\epsilon_{ijk}N_{\mu}^{i}
  N_{\nu}^{j}S^{k},\nn\\
&&X_{\mu}=Z_{\mu}-\half \epsilon_{ijk}\Gamma _{\mu}^{ij}S^{k},\nn
\end{eqnarray}
they generate Lorentz transformations of a coordinate frame
$(N_{\mu}^{0},N_{\mu}^{k}e_{i}^{k})$, by which the configuration
is decomposed with scalar coefficients. Thus, $M_{\mu\nu}$ generate
``rigid'' Lorentz transformations of the world sheet,
not changing its parametrization. Lorentz generators are
in involution with constraints: $\{M_{\mu\nu},\chi_{0,i}\}=0$.

Lorentz generators are simple functions of variables $(Z,P,\bf S)$,
which in our approach are independent, i.e. their quantum commutators
are postulated directly from Poisson brackets.
As a result, in quantum mechanics the commutators
$[M_{\mu\nu},M_{\rho\s}],\ [M_{\mu\nu},Q_{\rho}]$
{\it are anomaly free}. This can be proven by direct calculation,
done in \cite{slstring}.

The algebra of constraints $\chi_{i}$ in quantization acquires
the same anomaly that earlier was in Lorentz group. 
Thus, our current result is just a transfer of anomaly
from Lorentz group to the algebra of constraints.
However, now we have more freedom to solve the problem,
because we can impose additional gauges, excluding anomalous
component in the algebra of constraints. Gauges relate
the position of gauge axis with other dynamical vectors
in the system.

\paragraph*{Gauge 1:} let's direct $\es{3}\uparrow\uparrow\sbo\ 
\Leftrightarrow\ S_{1}=0,\ S_{2}=0$. These gauges are in week
involution with $\chi_{0,3}$ , and are the gauges only for $\chi_{1,2}$.
They are equivalent to 2nd class constraints onto oscillator
variables: $A^{\pm}=0,\ \{A^{+},A^{-}\}=2iS\neq0$.
The reduction of oscillator symplectic form onto the surface of 
these constraints leads to a form of complicated structure.
This structure becomes simple for a definite restricted class
of configurations.

\begin{figure}\label{f6}
\begin{center}
\parbox{6.4cm}{~\epsfysize=4cm\epsfxsize=6.4cm\epsffile{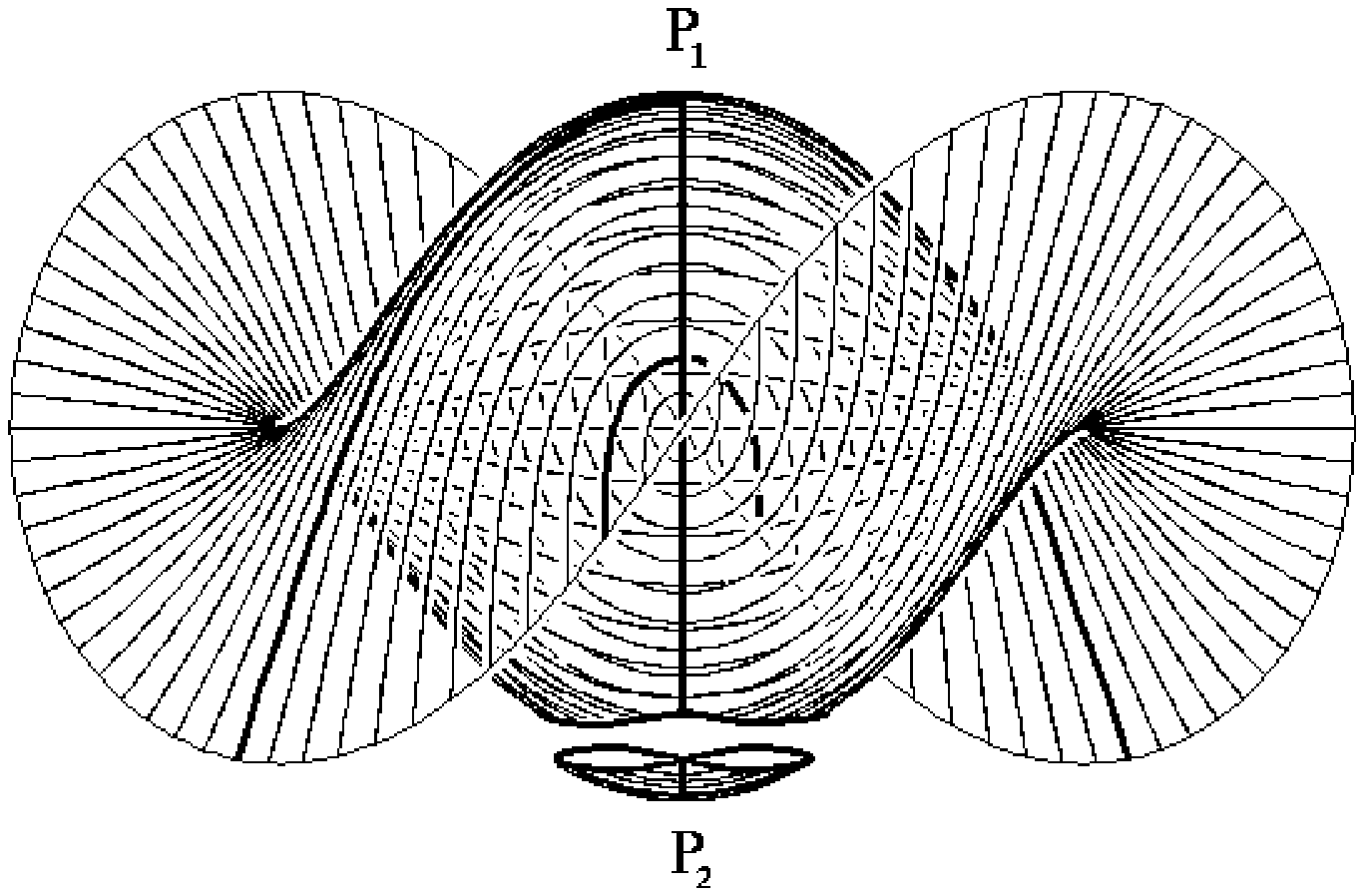}}\quad
\parbox{6cm}{\fignum Supporting curve and the world sheet with axial
symmetry of order 2 (both are projected to CMF).
Computer generated image \cite{vis}.}
\end{center}
\end{figure}

\paragraph*{Configurations with axial symmetry.}
Let's consider supporting curves, whose projection to CMF has
axial symmetry of order 2 (see \fref{f6}).

This restriction is equivalent to annulation of all even oscillator
variables: $a_{2n}=0,\ a_{2n}^{*}=0,\ n\in\Z$. 
On the surface of this restriction the constraints
$A^{\pm}=0$ are satisfied {\it identically}, 
because all terms in the sum $\sum \frac{1}{k}a_{k}a_{n}a_{k+n}^{*}$
vanish (if $k,n$ are odd, then $k+n$ is even). Reduction on the surface
of all constraints will give Poisson brackets:
\begin{eqnarray}
&&    \{Z_{\mu},P_{\nu}\}=g_{\mu\nu},\ \{a_{k},a_{n}^{*}\}=ik\delta_{kn},
\label{Pb2}\\
&&\{S^{i},S^{j}\}=-\epsilon^{ijk}S^{k},\
\{S^{i},e_{1}^{j}\}=-\epsilon^{ijk}e_{1}^{k},\nn
\end{eqnarray}
the difference from the general case (\ref{Pb1}) is that indices of oscillator variables
here are odd, and $({\bf S},{\bf e}_{1})$ represents the mechanics
of the rotator: ${\bf Se}_{1}=0,\ ({\bf e}_{1})^{2}=1$
(instead of the top in general mechanics). Remaining constraints:
$\chi_{0}=P^{2}/2\pi-L_{0},\ \chi_{3}=S-A_{3}$ are of the 1st class.
As in general mechanics, $\chi_{0}$ generates the evolution
$Q(\s)\to Q(\s+\tau)$ and $\chi_{3}$ does not change the configuration:
\hbox{$\{\chi_{3},Q(\s)\}=0$.}

\begin{figure}\label{f7}
\begin{center}
~\epsfxsize=5.3cm\epsffile{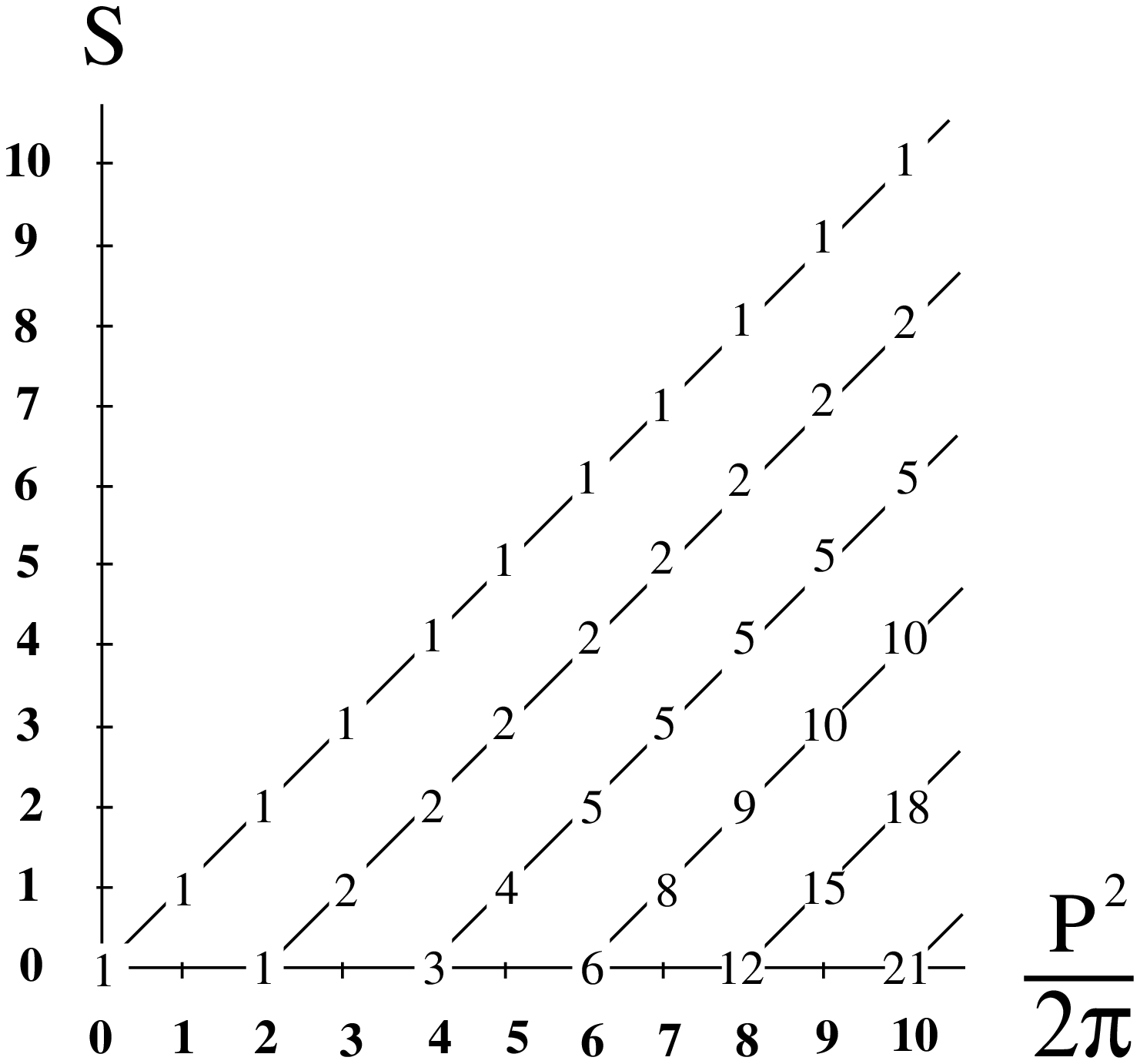}\quad
~\epsfxsize=5.3cm\epsffile{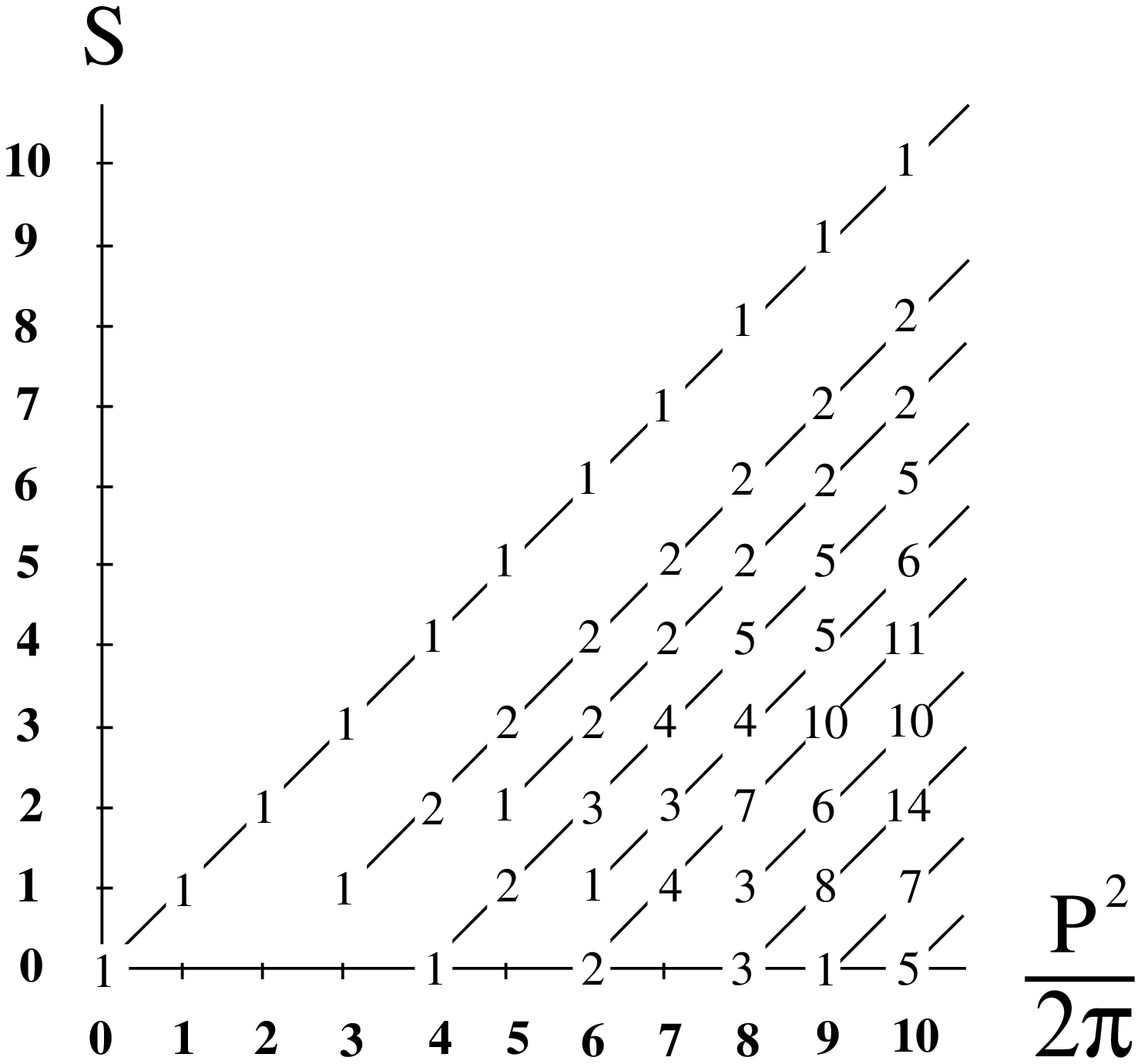}

\fignum On the left: spectrum of axially symmetrical configurations; 
on the right: spectrum of general configurations in Gupta-Bleuler's
approach; ($\delta\to+0$).
\end{center}
\end{figure}

\vspace{-3mm}
\paragraph*{Quantization} $\!\!\!$of this mechanics is straightforward.
Canonical operators
\begin{eqnarray}
&&[Z_{\mu},P_{\nu}]=-ig_{\mu\nu},  \  [a_{k},a_{n}^{+}]=k\delta_{kn},\ 
k,n~\mbox{odd},\nn\\ 
&&[S^{i},S^{j}]=i\epsilon^{ijk}S^{k},  \ 
    [S^{i},e_{1}^{j}]=i\epsilon^{ijk}e_{1}^{k},\
{\bf Se}_{1}=0,\ ({\bf e}_{1})^{2}=1\nn
\end{eqnarray}

\noindent
can be realized in a direct product of Fock space 
(with a vacuum \hbox{$a_{k}|0\ra>=0,\ k>0,$}
$a_{k}^{+}|0\ra>=0,\ k<0$)\fnm{a}\fnt{a}{
Such Fock space is positively defined,
and occupation number operators $n_{k}=\ :a_{k}^{+}a_{k}:/|k|$
take values: 0,1,2...} onto
the space of functions $\Psi(P,{\bf e}_{1})$, with definition of operators
$Z=-i\df/\df P,\ {\bf S}=-i{\bf e}_{1}\times\df/\df{\bf e}_{1}$.
Physical subspace is defined by 2 constraints
$\left({\textstyle{{P^{2}}\over{2\pi}}}-\!\!\!\sum\limits_{odd~k}
|k|n_{k}-\delta\right)|\Psi\ra>=0,
\ \left(S-\!\!\!\sum\limits_{odd~k}\mbox{sign~}
k\cdot n_{k}\right)|\Psi\ra>=0$ (here $\delta>0$ is arbitrary
c-number). Spin-mass spectrum of this mechanics is shown on 
\fref{f7}. It's also possible to construct the operator 
$Q_{\mu}(\s)$, which satisfies all necessary requirements:
is finite, Hermitian and commutators $[Q(\s),\chi_{0,3}]$
repeat classical Poisson brackets (see \cite{ax} for details).

\vspace{2mm}

\vspace{2mm}
\noindent\underline{{\it Remark:}} 
general configurations in Gupta-Bleuler's approach. 
We considered the mechanics of restricted configurations. 
For general configurations
one should deal with a complicated mechanics, appearing
in the reduction on the 2nd class constraints $A^{\pm}=0$.
However, it's possible to construct
self-consistent quantum theory by methods, deviating
from standard Dirac's procedure. For example, let's
replace two 2nd class constraints by a single constraint:
$A^{\pm}=0\ (A_{1,2}=0)\ \rightarrow\ A^{-}A^{+}=A_{1}^{2}+A_{2}^{2}=0$.
In quantum theory we will have $A^{-}A^{+}|\Psi\ra>=0\ \Rightarrow\ 
\la<\Psi|A^{-}A^{+}|\Psi\ra>=0$, and
{\it in positively defined space} we are using this is equivalent
to $A^{+}|\Psi\ra>=0$ -- the result coincides with Gupta-Bleuler's imposition
of Hermitian conjugated constraints. Additionally we have 
the constraints $(S-A_{3})|\Psi\ra>=0,\ 
(P^{2}/2\pi-L_{0}-\delta)|\Psi\ra>=0$, the whole set of constraints belongs 
to the 1st class. Moreover, because $L_{0}=\sum|k|n_{k}$,
at any fixed value of $P^{2}/2\pi$ the mass shell condition
defines {\it finite-dimensional} subspace, where other constraints
act as matrices of finite size. It's easy to resolve the correspondent
linear systems (see Appendix~3) and find the spectrum, shown on 
\fref{f7}, right.

It's necessary to understand, that the described method
is {\it different} from Dirac's approach,
where imposition of constraint $\chi|\Psi\ra>=0$
should imply, that physical states are invariant
under the action of a gauge transformations, generated by 
$\chi$. In our case 
the replacement of two 2nd class constraints
by their sum of squares gives {\it a degenerate} constraint,
which generates no transformations. 

This method 
of constraints imposition has some physical ground.
Let's consider a simplest example of 2nd class constraints:
$x=0,\ p=0$. The Dirac's reduction completely eliminates
these 2 degrees of freedom. If we replace these constraints
by a single one: $x^{2}+p^{2}=a^{*}a=0,$ after an appropriate choice of
quantum ordering we will have a constraint $a^{+}a|\Psi\ra>=0$,
defining the {\it vacuum} state, where $x,p$ are distributed
by Gaussian function with the dispersion $\hbar$. Taking $\hbar\to0$,
we will have the same situation, as in classical theory.

Realization of {\it 2nd class} constraints 
by a single constraint a'la Gupta-Bleuler
is widespread in literature (example is
the method of string quantization, which was proposed by Rohrlich 
\cite{Rohrlich} and then was extensively used for a 
construction of hadrons' models, 
see \cite{Barbashov}).
We, however, want to use purely Dirac's methods and now will
try to quantize the theory in another gauge.

\vspace{-3mm}
\paragraph*{Gauge 2, general configurations:}
let's direct ${\bf e_{3}}\perp\sbo\ \Leftrightarrow\ S_{3}=0$.
Replacing the constraints $S^{\pm}-A^{\pm}=0$ by equivalent 
combination $|S^{+}|-|A^{+}|=0,\ \mbox{arg~}S^{+}-\mbox{arg~}A^{+}=0$,
we see that $S_{3}=0$ is a gauge only to the last constraint
(it generates phase rotations of $S^{+}$), and is in involution with others.

Reduction leads to a mechanics, different from general one (\ref{Pb1})
by a replacement: the top $\e,{\bf S}\ \to\ $ the rotator 
${\bf e}_{3},{\bf S}$. Mechanics includes three 1st class
constraints: $P^{2}/2\pi-L_{0}=0,\ A_{3}=0,\ S-\sqrt{A^{+}A^{-}}=0$.
First two constraints generate phase rotations of oscillator variables:
$a_{n}\to a_{n}e^{-in\tau},\ a_{n}\to a_{n}e^{-i\tau}$,
and $A^{+}A^{-}$ is a polynomial of $a_{n},a_{n}^{*}$ with such structure,
that it is conserved in these phase rotations. Because this property
can be easily preserved in quantization, the quantum algebra
of constraints will be free of anomalies. Configuration 
$a^{\alpha}(\s)$ has a form:

\vspace{-3mm}
\begin{eqnarray}
a^{\alpha}(\sigma) &=&  \left(
 \frac{\pi}{2\sqp}\left(\frac{P^{2}}
  {\pi^{2}}+|a(\sigma)|^{2}\right),
{\textstyle{{1}\over{2S}}}
(a(\s)A^{+}{\bf n}+\mbox{~c.c.~})\right.\label{conf2}\\
&&+\left.\frac{\pi}{2\sqp}\left(\frac{P^{2}}{\pi^{2}}-|a(\sigma)|^{2}
\right){\bf e}_{3} \right),\nn
\end{eqnarray}

\noindent
where ${\bf n}=({\bf S}- i{\bf S}\times{\bf e}_{3})/S$. 
As earlier, mass shell condition generates the evolution
$Q(\s)\to Q(\s+\tau)$ and $\{A_{3},Q(\s)\}=0$
($A_{3}$ rotates the phases of $a(\s)$ and $A^{+}$ in opposite
directions and conserve $a(\s)A^{+}$). The constraint 
$S-\sqrt{A^{+}A^{-}}=0$ generates the
rotations of gauge axis about spin vector, and correspondent
reparametrizations of the supporting curve.

\vspace{2mm}\noindent
{\it P-reflection operation} can be defined in this mechanics,
consisting of two factors: 

\noindent$\Pi:$\quad
reflection of the supporting curve w.r.t. a plane, perpendicular to 
${\bf S}$, 
which is performed by a replacement $a_{n}\to a_{-n}^{*}\ \Rightarrow\
a(\s)\to a^{*}(\s),\ A^{+}\to-A^{-}$;

\noindent$R:$\quad
rotation of the supporting curve about spin with angle $\pi$:
${\bf e}_{3}\to-{\bf e}_{3},\ {\bf S}=Const\ \Rightarrow\
{\bf n}\to{\bf n}^{*}$;

\noindent so that ${\bf a}(\s)\to-{\bf a}(\s)$.

\vspace{2mm}
\noindent\underline{{\it Remark}}: in gauge 1 the gauge axis was directed
along the spin, which is not changed in P-reflection.
Thus, the gauge axis is not changed also. As a result,
P-reflection changes a position of the supporting curve w.r.t. gauge axis
and is followed by reparametrization. This makes problematic
the definition of P-reflection in gauge~1.

\vspace{-3mm}
\paragraph*{Quantization.} Canonical operators:

\vspace{-5mm}
$$[Z_{\mu},P_{\nu}]=-ig_{\mu\nu},  \ [a_{k},a_{n}^{+}]=k\delta_{kn},\ 
k,n\in\Z/\{0\},\
[S^{i},S^{j}]=i\epsilon^{ijk}S^{k},  \
    [S^{i},e_{3}^{j}]=i\epsilon^{ijk}e_{3}^{k},$$

\noindent where ${\bf Se}_{3}=0,\ ({\bf e}_{3})^{2}=1$.
Realization is analogous to presented above. Operators $A^{\pm}$,
defined by the same polynomial expressions as in (\ref{Apol}),
have no ordering ambiguities. Three constraints:
$$(P^{2}/2\pi-L_{0}-\delta)|\Psi\ra>=0,\quad A_{3}|\Psi\ra>=0,\quad
\left(S-\sqrt{(A^{+}A^{-}+A^{-}A^{+})/2}~\right)|\Psi\ra>=0$$

\noindent belong to the 1st class. Symmetric ordering under the square root
in the last constraint was chosen, because it commutes with
P-reflection, defined as follows:

\vspace{-5mm}
\begin{eqnarray}
&&P=R\Pi,\ R=e^{i\pi S},\ \Pi=\{a_{k}\to a_{-k}^{+}\}=
\{|n_{k}\ra>\to|n_{-k}\ra>\}\nn\\
&& \Rightarrow\
\Pi A^{+}\Pi=-A^{-},\ \Pi A^{+}A^{-}\Pi=A^{-}A^{+}.\nn
\end{eqnarray}

Again, at fixed $P^{2}/2\pi$ the mass shell condition
defines finite-dimensional subspace, where other constraints
act as finite matrices. It's easy to solve the eigenvalue
problem for these matrices (see Appendix~3).

\begin{figure}\label{f14}
\begin{center}
~\epsfysize=5.3cm\epsfxsize=5.3cm\epsffile{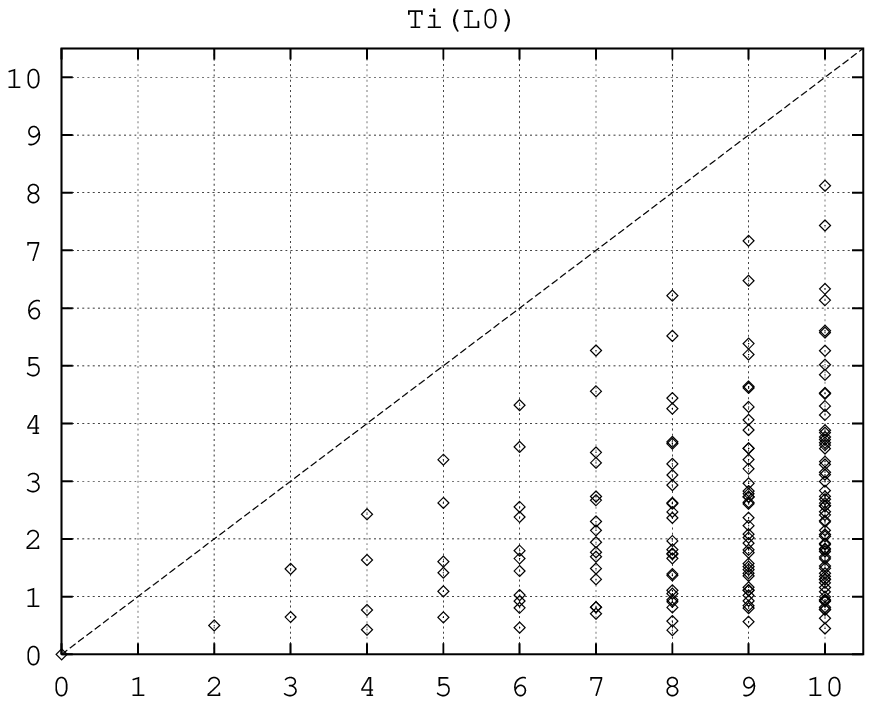}\quad
~\epsfysize=5.3cm\epsfxsize=5.3cm\epsffile{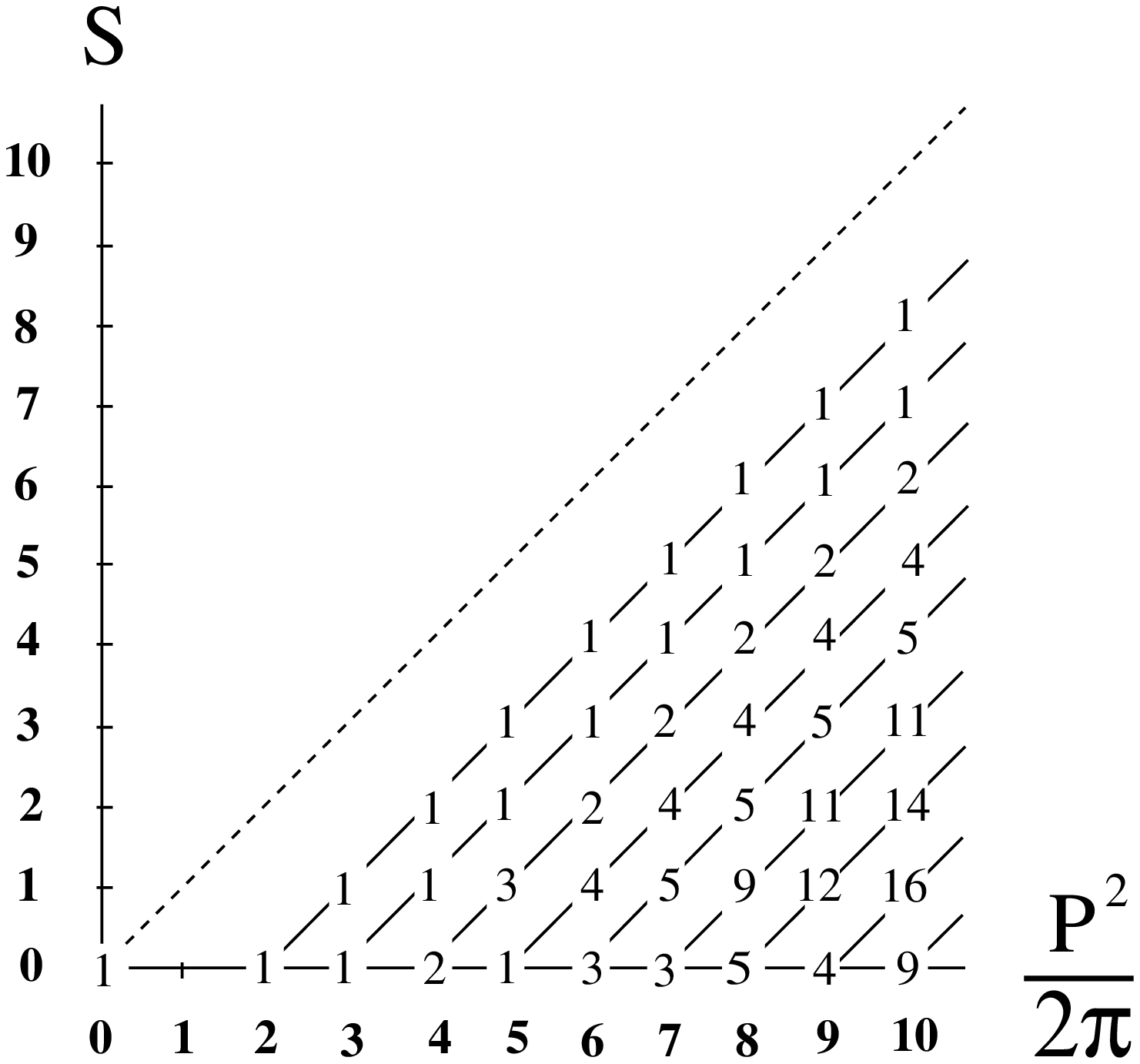}

\fignum On the left: spectrum $T(L_{0})$; 
on the right: integer part of this spectrum.
\end{center}
\end{figure}

The main obstacle now is that operators $A^{\pm}$ 
still have the anomaly in commutator. If they would not have
anomaly, the definition of the square root
$T=\sqrt{(A^{+}A^{-}+A^{-}A^{+})/2}$, analogous to
$S=\sqrt{{\bf S}^{2}+1/4}-1/2\ \Leftrightarrow\ {\bf S}^{2}=S(S+1)$
will give integer spectrum for $T$. But now $A^{\pm}$ have anomaly
and do not represent the rotation group. The spectrum of $T$
is not integer (even in S-like definition), see \fref{f14}.
Due to the constraint $S-T=0,\ S\in\Z,\ T\notin\Z$ 
the theory becomes empty.

This problem (anomaly in spectrum) is absolutely different
from the usual one (anomaly in commutator). The following
solution can be proposed.

Generally we can add to $T$ any operator, which
commutes with all constraints, and whose contribution
is classically vanishing (e.g. eigenvalues of $\delta T$ are 
bounded by Planck's constant: $|\delta T_{i}|<Const\cdot\hbar$).
The corrected variable will have the same classical limit as $T$. 
With the aid of these corrections we can {\it deform} the spectrum of $T$
to integer values. There is an infinite number of possible deformations, 
the simplest one: shift eigenvalues of $T$ to the closest integer value below.
Because we don't change the eigenvectors of $T$, 
the constraints remain to be of the 1st class
($T$ acts in the subspaces, defined by other constraints).
Redefinition changes $T$ by operator, whose eigenvalues
are restricted between 0 and 1, or when we restore
Planck's constant -- between 0 and $\hbar$. Therefore, 
the corrected operator $T$ has the same classical limit\fnm{b}\fnt{b}{
Quasi-classical approximation also gives integer values for $T$
spectrum ($T$ is action-type variable, generating
$2\pi$-periodical evolution).}. 
After the redefinition $T\to[T]$ we have spin-mass spectrum,
shown on \fref{f14}, right.

\vspace{1mm}

\vspace{2mm}
\noindent\underline{{\it Remark}}:
Leading Regge trajectory (starting from the vacuum state
$L_{0}=0$) is seemingly absent in the spectrum.
This effect, of course, can be explained as anomaly-induced
nonlinearity of the leading trajectory $T_{i}(L_{0})$
(\fref{f14}, left). However, there are arguments,
why the leading trajectory really should be absent in this
approach, see Appendix~4. 

\paragraph*{Vertex operators.}
$\!\!\!$The operator $a^{\alpha}(\sigma)$ has ambiguities in ordering.
A possible definition is:
\begin {eqnarray}
a^{\alpha}(\sigma) &=& \left(
 \frac{1}{\sqp}\left(\frac{P^{2}}
  {\pi}+\sum_{n\neq 0}L_{n}e^{-in\sigma}  \right),
\ \left(a(\s)A^{+}{\textstyle{{1}\over{2S+1}}}
{\bf n}+\mbox{~h.c.~}\right)\right.\label{asimpl}\\
&&\left.-\frac{1}{\sqp}\sum_{n\neq 0}L_{n}e^{-in\sigma}
    {\bf e}_{3}\right).\nn
\end{eqnarray}
Here $L_{n}=\sum a_{k}a^{+}_{k-n}$ and
${\bf n}={\textstyle{{1}\over{S+1/2}}}({\bf S}-i{\bf S}\times{\bf e}_{3}
-{\bf e}_{3}/2)$, see \cite{em}.
Configuration operator 
is finite\fnm{c}\fnt{c}{Matrix elements of this operator 
between states with finite number of excited modes are finite.}, 
Hermitian and satisfies
the relations: $[a^{\alpha}(\s),L_{0}]=ia^{\alpha}(\s)',\ 
[a^{\alpha}(\s),A_{3}]=0$.

Anomaly can appear in commutator 
$[a^{\alpha}(\s),S-T]$\fnm{d}\fnt{d}{because 
there are anomalous operators under
the square root in $T$, and because we perform redefinitions $T\to[T]$.}.
In classical mechanics the Poisson bracket $\{a^{\alpha}(\s),S-T\}$
is a complicated non-linear expression of independent variables, 
and we have the ordering ambiguity in definition of correspondent
quantum expression. In principle, one can try to preserve this 
commutation relation by an appropriate choice of ordering procedure,
or by classically vanishing corrections of simplest definition (\ref{asimpl}).
However, now we will show, that anomaly in this commutator is not
crucial for the theory, because in spite of this anomaly, we are able 
to construct the vertex operators, acting in the physical subspace.

Operators $a^{\alpha}(\sigma)$ themselves do not act in the physical subspace: 
$a^{\alpha}(\s)|phys\ra>\not\subset|phys\ra>$, due to 
$[a^{\alpha}(\s),L_{0}]\neq0$. Vertex operators are their special combinations.
For example, emission of photons from the charged ends of 
string is described by operator \cite{em}:
$V^{\alpha}=\int_{0}^{2\pi}\frac{d\s}{2\pi} :a^{\alpha}(\s)e^{ikQ(\s)}:$
If initial and final states satisfy the mass shell condition
$P^{2}/2\pi-(P-k)^{2}/2\pi\in\Z$,
this operator acts in the physical subspace. Indeed,
acting by this operator on the state:
$$V^{\alpha}|\frac{P^{2}}{2\pi},{\chi_{0}=0}\ra>=
\int_{0}^{2\pi}\frac{d\s}{2\pi} e^{i\chi_{0}\s}:a^{\alpha}(0)e^{ikQ(0)}:
e^{-i\chi_{0}\s}|\frac{P^{2}}{2\pi},{\chi_{0}=0}\ra>$$
(here we explicitly extract 
$\s$-dependence by embracing evolution operators),
we will have the right evolution operator equal to unity,
and $\int_{0}^{2\pi}\frac{d\s}{2\pi} e^{i\chi_{0}\s}$ becomes
projector to the space with $\chi_{0}=0$. As a result, we obtain physical 
state $|\frac{(P-k)^{2}}{2\pi},{\chi_{0}=0}\ra>$ ($e^{ikQ(0)}$
includes $e^{ikZ}$ operator, shifting $P\to P-k$).

In analogous way we can define the vertex operator, commuting with the
constraint $\Lambda=S-T$: $\tilde V^{\alpha}=
\int_{0}^{2\pi}\frac{d\tau}{2\pi} V^{\alpha}(\tau),\ 
V^{\alpha}(\tau)=e^{i\Lambda\tau}V^{\alpha}e^{-i\Lambda\tau}$.
The $\tau$-dependence of $V^{\alpha}(\tau)$ corresponds 
to the rotations of gauge axis about spin (transformations,
generated by $\Lambda$), and we average $V^{\alpha}$
by these rotations. However, in classical mechanics $V^{\alpha}$
is parametric invariant (constant on $\tau$), consequently, 
$V^{\alpha}(\tau)=V^{\alpha}+f(\tau)$, where all $\tau$-dependent terms 
are $f(\tau)=O(\hbar)$. Therefore, the constructed vertex operator 
$\tilde V^{\alpha}$ classically corresponds to the same variable
$V^{\alpha}$ (changes, performed by insertion of evolution operators
and averaging are $O(\hbar)$). Now it acts in the physical space.
The proof is analogous: $\tilde V^{\alpha}|{\Lambda=0}\ra>=
\int_{0}^{2\pi}\frac{d\tau}{2\pi} e^{i\Lambda\tau} V^{\alpha}
e^{-i\Lambda\tau}|{\Lambda=0}\ra>$,
in the presence of the physical state $e^{-i\Lambda\tau}=1$, and 
$\int_{0}^{2\pi}\frac{d\tau}{2\pi} e^{i\Lambda\tau}$ becomes a projector 
to $\Lambda=0$ space. 

Also there is an obvious identity 
$\la<{phys'}|\tilde V^{\alpha}|{phys}\ra>=
\la<{phys'}|:a^{\alpha}(0)e^{ikQ(0)}:|{phys}\ra>$.
Thus, in practical calculations it's sufficient to find
matrix elements of operator $:a^{\alpha}(0)e^{ikQ(0)}:$
between the physical states. Non-physical states do not appear
in these calculations.

\vspace{-3mm}
\paragraph*{Conclusion.} 
We have constructed a quantum theory, which acts in 4-dimensional space-time,
does not have anomalies in Lorentz group and algebra of constraints,
and in classical limit represents Nambu-Goto string theory.
We considered 3 variants of this theory:

\begin{itemize}

\vspace{-2mm}\item[1.] 
quantization of axially symmetrical world sheets
(\fref{f7}, left);

\vspace{-2mm}\item[2.] 
quantization of general world sheets by Gupta-Bleuler's procedure
(\fref{f7}, right);

\vspace{-2mm}\item[3.] 
quantization of general world sheets by Dirac's procedure
(\fref{f14}, right).
\end{itemize}
In the first two cases the quantum theory has no intrinsic
difficulties. In the third case there is anomaly in spectrum
of an operator $T$, entering into one of the constraints. 
To obtain non-empty theory, we should perform classically vanishing
corrections of this operator. These corrections are ambiguous,
only one variant was considered. Independently on the definition
of $T$, vertex operators can be constructed, acting in the 
physical subspace. 

\vspace{-3mm}
\paragraph*{Acknowledgment.}
Author is indebted to Prof. George P. Pronko for valuable discussions.
The work has been partially supported by \hbox{INTAS~96-0778} grant.

\baselineskip=0.4\normalbaselineskip\footnotesize

\paragraph*{Appendix 1: Symplectic structure of the phase space 
\cite{slstring,Arnold}.}\quad 

\vspace{1mm}\noindent
In modern formulation of Hamiltonian mechanics the phase space
is defined as a smooth manifold, endowed by a closed non-degenerate 
differential 2-form $\Omega = {{1}\over{2}}\omega_{ij}dX^{i}\wedge dX^{j}$ 
(in some local coordinates
 $X^{i},\  i=1,\ldots,2n$). Poisson brackets are defined by the form as
$\{X^{i},X^{j}\}=\omega^{ij}$, where $\|\omega^{ij}\|$ is inverse to 
$\|\omega_{ij}\|$: $\omega_{ij}\omega^{jk}=\delta_{i}^{k}$.

Let's consider a surface in the phase space, given by the 2nd class 
constraints:
 $\chi_{\alpha}(X)=0\ (\alpha =1,\ldots,r),$ $det\|\{\chi_{\alpha},
\chi_{\beta}\}\|\neq 0 $. Reduction on this surface consists in the 
substitution of
its explicit parametrization $X^{i}=X^{i}(u^{a})\ (a=1,\ldots,2n-r)$ into the 
form: 
$$\Omega = {{1}\over{2}}\Omega_{ab}du^{a}\wedge du^{b},\quad
 \Omega_{ab}={{\partial X^{i}}\over{\partial u^{a}}}\omega_{ij}
 {{\partial X^{j}}\over{\partial u^{b}}},\quad
 det\|\Omega_{ab}\|\neq 0. $$
Matrix $\|\Omega^{ab}\|$ , inverse to 
$\|\Omega_{ab}\|$, defines Poisson brackets on the surface:
 $\{u^{a},u^{b}\}=\Omega^{ab}$.

This method is equivalent to commonly used Dirac brackets' formalism.
Sometimes it is convenient to combine both methods: some of the constraints
$\chi_{\alpha}(X)$ are imposed as above, then Dirac brackets on the remaining
constraints $\psi_{n}(u)$ are calculated by definition:
$$ \{u^{a},u^{b}\}^{D}=\{u^{a},u^{b}\} - \{u^{a},\psi_{n}\}\Pi^{nm}
\{\psi_{m},u^{b}\},$$
where $\|\Pi^{nm}\|$ is inverse to $\|\Pi_{nm}\|$: $\Pi_{nm}=
\{\psi_{n},\psi_{m}\}$.

In string theory canonical Poisson brackets 
$\{x_{\mu}(\s),p_{\nu}(\tilde\s)\}=g_{\mu\nu}\delta(\s-\tilde\s)$
correspond to symplectic form $\Omega=\int_{0}^{\pi}d\s\;
\delta p_{\mu}(\s)\wedge \delta x_{\mu}(\s)$.
It can be transformed to the form (\ref{orig})
by a substitution of expressions for $x_{\mu}(\s),p_{\mu}(\s)$
in terms of $Q_{\mu}(\s)$; and then to the form (\ref{Olcg}) by substitution
of light cone parametrization for $Q_{\mu}(\s)$.

Spin part of the form 
$\Omega_{S}=\half d{\bf e}_{i}\wedge d({\bf S}\times{\bf e}_{i})$
corresponds to Poisson brackets (\ref{Pb1}). To prove this,
at first invert coefficient matrix of the form, ignoring
orthonormality constraints ${\bf e}_{i}{\bf e}_{j}=\delta_{ij}$,
then calculate Dirac's brackets on the surface of these constraints.

Considering {\it gauge 1}, we substitute ${\bf e}_{3}={\bf S}/S,\
{\bf e}_{2}={\bf e}_{3}\times{\bf e}_{1}$ into the form and find
$\Omega_{S}=d{\bf e}_{1}\wedge d({\bf S}\times{\bf e}_{1})$.
Correspondence of this form to Poisson brackets (\ref{Pb2})
can be proven analogously.

For {\it gauge 2} we use the following property
of symplectic form.

\vspace{3mm}
\hfill\parbox[t]{3cm}{\begin{figure}\label{f18}
~\epsfysize=5.3cm\epsfxsize=3cm\epsffile{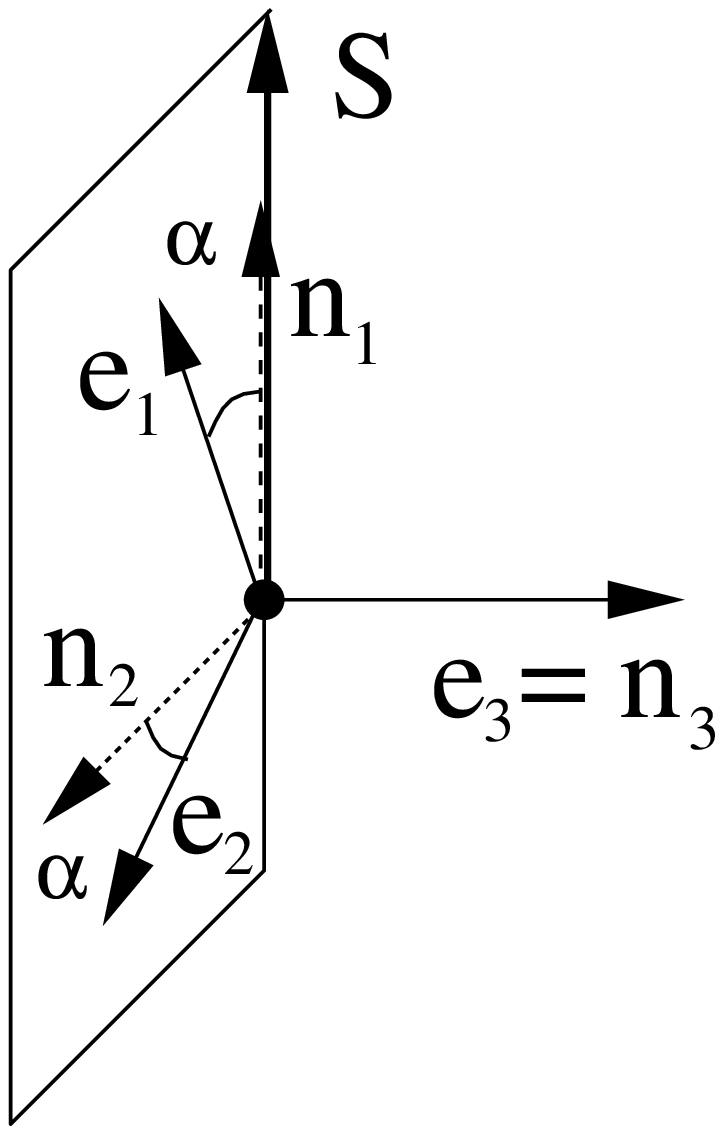}

\vspace{3mm}
\parbox{3cm}
{\fignum \\ Rotation \\ of the basis \\ for gauge 2.}
\end{figure}
}\quad\quad
\parbox[t]{8cm}{
{\it Lemma.} Let ${\bf e}_{i}=R_{ij}^{(k)}{\bf n}_{j},$ 
where $R_{ij}^{(k)}(\alpha)$ is a matrix of rotation
about axis ${\bf n}_{k}:\ R^{(k)}=\exp\alpha\hat r^{(k)},\
\hat r_{ij}^{(k)}=\epsilon_{ijk}$. Then 
$\Omega_{S}=\half d{\bf n}_{i}\wedge d({\bf S}\times{\bf n}_{i})
+d\alpha\wedge dS_{k}$, where $S_{k}={\bf n}_{k}{\bf S}$
is a projection of spin vector to rotation axis.

\vspace{3mm}\noindent
{\it Proof.}\quad$
\Omega_{S}=\half d{\bf n}_{i}\wedge d({\bf S}\times{\bf n}_{i})
-\half({\bf S}\times{\bf n}_{j},{\bf n}_{k})dR_{ij}^{(k)}\wedge dR_{ik}^{(k)}+
\half\epsilon_{kjl}\; R_{ik}^{(k)}dR_{ij}^{(k)}\wedge d({\bf n}_{l}{\bf S}).$
The second term here is proportional $d\alpha\wedge d\alpha=0$.
Using the identity $(R^{(k)})^{T}dR^{(k)}=\hat r^{(k)}d\alpha$ 
in the third term, transform it to the form $d\alpha\wedge dS_{k}$.
Lemma is proven.

\vspace{3mm}
Let's consider the gauge 2: ${\bf e}_{3}\perp{\bf S}$ (\fref{f18}),
and rotate the basis about ${\bf e}_{3}$ to align the first axis
along ${\bf S}$: ${\bf e}_{i}=R_{ij}^{(3)}(\alpha){\bf n}_{j};\
{\bf n}_{3}={\bf e}_{3},\ {\bf n}_{1}={\bf S}/S,\ 
{\bf n}_{2}={\bf n}_{3}\times{\bf n}_{1};\ \cos\alpha=S_{1}/S,\
\sin\alpha=-S_{2}/S\ \Rightarrow\ \alpha=\mbox{arg~}S^{-}=\mbox{arg~}A^{-}$.
Because in this rotation $S_{3}=0$, the additional term 
$d\alpha\wedge dS_{3}=0$. The resulting form can be 
transformed to $\Omega_{S}=d{\bf n}_{3}\wedge d({\bf S}\times{\bf n}_{3}),\
{\bf n}_{3}={\bf e}_{3}$: the top $({\bf e}_{i},{\bf S})$ is replaced by
the rotator $({\bf e}_{3},{\bf S})$.
Replacement ${\bf e}_{i}\to{\bf n}_{i}$ in the expression for
configuration (\ref{conf1}) transforms it to (\ref{conf2}).
}

\paragraph*{Appendix 2. Geometrical reconstruction of the world sheets}\quad 

\vspace{1mm}\noindent
Properties 1,2 of supporting curve follow from its definition.
Property 3 follows from 4 and 2. Let's prove the property~4:
\begin{eqnarray}
&&x_{\mu}(\s,\tau)=(Q_{\mu}(\s_{1})+Q_{\mu}(\s_{2}))/2,\quad 
\s_{1,2}=\tau\pm\s.\label{sheet}
\end{eqnarray}
This formula was obtained in \cite{zone} by direct solution of Hamiltonian 
equations in $Q$-representation. Here we will reproduce the proof 
of this formula in oscillator representation.

Coordinates and momenta of the string are defined by 
expressions\fnm{e}\fnt{e}{
See e.g.\cite{Brink}. Difference of notations: $a_{\mu}^{n}$ 
in our work corresponds to $i\sqrt{n}a_{\mu}^{n*}$ in \cite{Brink} ($n>0$).}
\begin{eqnarray}
&&x_{\mu}(\sigma)=X_{\mu}+{\textstyle{{1}\over{\sqrt{\pi}}}}
\sum\limits_{n\neq0}{\textstyle{{a_{\mu}^{n}}\over{in}}}\cos n\sigma,\quad
p_{\mu}(\sigma)={\textstyle{{1}\over{\sqrt{\pi}}}}
\sum\limits_{n}a_{\mu}^{n}\cos n\sigma,\nn
\end{eqnarray}
so that formulae (\ref{Qini0})-(\ref{Qosc}) are valid.
Poisson brackets for canonical variables:
\begin{eqnarray}
&&\{a_{\mu}^{n},a_{\nu}^{k}\}=in\;g_{\mu\nu}\;\delta^{k,-n}, 
\quad \{X_{\mu},P_{\nu}\}=g_{\mu\nu}.\nn
\end{eqnarray}
Hamiltonian of the system is an arbitrary linear combination of Virasoro
constraints: $H=\sum c^{k}L^{k}\  (L^{k}=\sum_{n} a_{\mu}^{n}a_{\mu}^{k-n},\
c^{k*}=c^{-k})$.
Coefficients $c^{k}$ influence only parametrization of the world sheet.
The choice $H=L^{0}$ corresponds to conformal parametrization (where
$(x'\pm\dot x)^{2}=0$). This Hamiltonian generates phase rotations
$a_{\mu}^{n}(\tau)=a_{\mu}^{n}(0)e^{in\tau}$ and shifts 
$X_{\mu}(\tau)=X_{\mu}(0)+(P_{\mu}/\pi)\tau $.
Using (\ref{Qosc}), we see that the evolution of function $Q_{\mu}(\sigma)$ 
is the shift of its argument: $Q_{\mu}(\sigma,\tau)
=Q_{\mu}(\tau+\sigma,0)$. Then, using (\ref{Qini}), we have the 
following evolution for coordinates and momenta:
\begin{eqnarray}
&&x_{\mu}(\sigma,\tau)=(Q_{\mu}(\tau+\sigma,0)+Q_{\mu}(\tau-\sigma,0))/2,\quad
p_{\mu}(\sigma,\tau)=(Q'_{\mu}(\tau+\sigma,0)+Q'_{\mu}(\tau-\sigma,0))/2.\nn
\end{eqnarray}
Introducing isotropic coordinates $\sigma_{1,2}=\tau\pm\sigma$,
obtain formula (\ref{sheet}).

\vspace{3mm}
\noindent\underline{{\it Remark:}}
Using Poisson brackets (\ref{orig}), 
we have $\{Q_{\mu}(\s),Q'^{2}(\tilde\s)/4\}=
\Delta(\s-\tilde\s)Q'_{\mu}(\s),$ and for 
$H=\int d\s\;F(\s)Q'^{2}(\s)/4:\quad 
\dot Q_{\mu}(\s)=\{Q_{\mu}(\s),H\}=F(\s)Q'_{\mu}(\s)$,
linear combinations of constraints generate shifts of points 
in tangent direction to the supporting curve, or equivalently --
reparametrizations of this curve. 

\noindent 1. Taking $F=1$, we will obtain the evolution 
$Q(\sigma)\to Q(\tau+\sigma)$ and formula (\ref{sheet}) again.

\noindent 2. Considering arbitrary $F$, we will see that
the reduced phase space of string 
(obtained in factorization of the phase space
by the action of gauge group) is actually a set of all possible supporting 
curves, which are considered as geometric images, without respect to their
parametrization (two different parametrizations of the curve correspond to
the same point of the reduced phase space). 
All physical observables in string theory are parametric invariants 
of supporting curve. The world sheet is also reconstructed by the supporting 
curve in parametrically invariant way, see \fref{f1}. 

\paragraph*{Appendix 3: Bases of physical subspaces.}\quad 

\vspace{3mm}
\noindent{\it Gauge 1, axially symmetrical configurations.}
Basis of physical subspace is given in Table~1.

\vspace{1mm}
\noindent\parbox{7cm}{
{\it Gauge 1, Gupta-Bleuler's approach.}
Fig.\ref{f20} shows the spectrum of operators $(L_{0},A_{3})$.
Due to the constraint $S-A_{3}=0,\ S\geq0$, we should consider
only the upper part of this spectrum: $A_{3}\geq0$.

Operator $A^{+}$ acts on the level $L_{0}=Const$ and raises $A_{3}$ by 1.
It annulates all states on leading trajectory $L_{0}=A_{3}$ 
(among them $|L_{0}=0,A_{3}=0\ra>$ and $|L_{0}=1,A_{3}=1\ra>$).
For the states with $0\leq A_{3}<L_{0}, L_{0}\geq2$ the multiplicity of states
on the level $N(L_{0},A_{3})$ has a property 
$N(L_{0},A_{3})\geq N(L_{0},A_{3}+1)$. The remaining part of consideration
is a proof (done by direct computation), that matrices
$A^{+}_{ij}=\la<L_{0},A_{3}+1,i|A^{+}|L_{0},A_{3},j\ra>$,
representing linear map $A^{+}:\ (L_{0},A_{3})\to(L_{0},A_{3}+1)$,
have maximal rank (equal to less dimension $N(L_{0},A_{3}+1)$).
Therefore, this linear map has a kernel with the dimension
$K(L_{0},A_{3})=N(L_{0},A_{3})-N(L_{0},A_{3}+1)$.
(Particularly, the linear map $(3,1)\to(3,2)$, marked by arrow on \fref{f20},
has exactly 1-dimensional kernel.) Computing $K(L_{0},A_{3})$
for the whole spectrum \fref{f20}, we obtain the spectrum \fref{f7}, right.
Then we solve the linear equations $A^{+}_{ij}\Psi_{j}=0$,
and place the result in Table~2.

}\quad
\parbox{6cm}{

\begin{figure}\label{f20}
\begin{center}
~\epsfysize=7.5cm\epsfxsize=5.3cm\epsffile{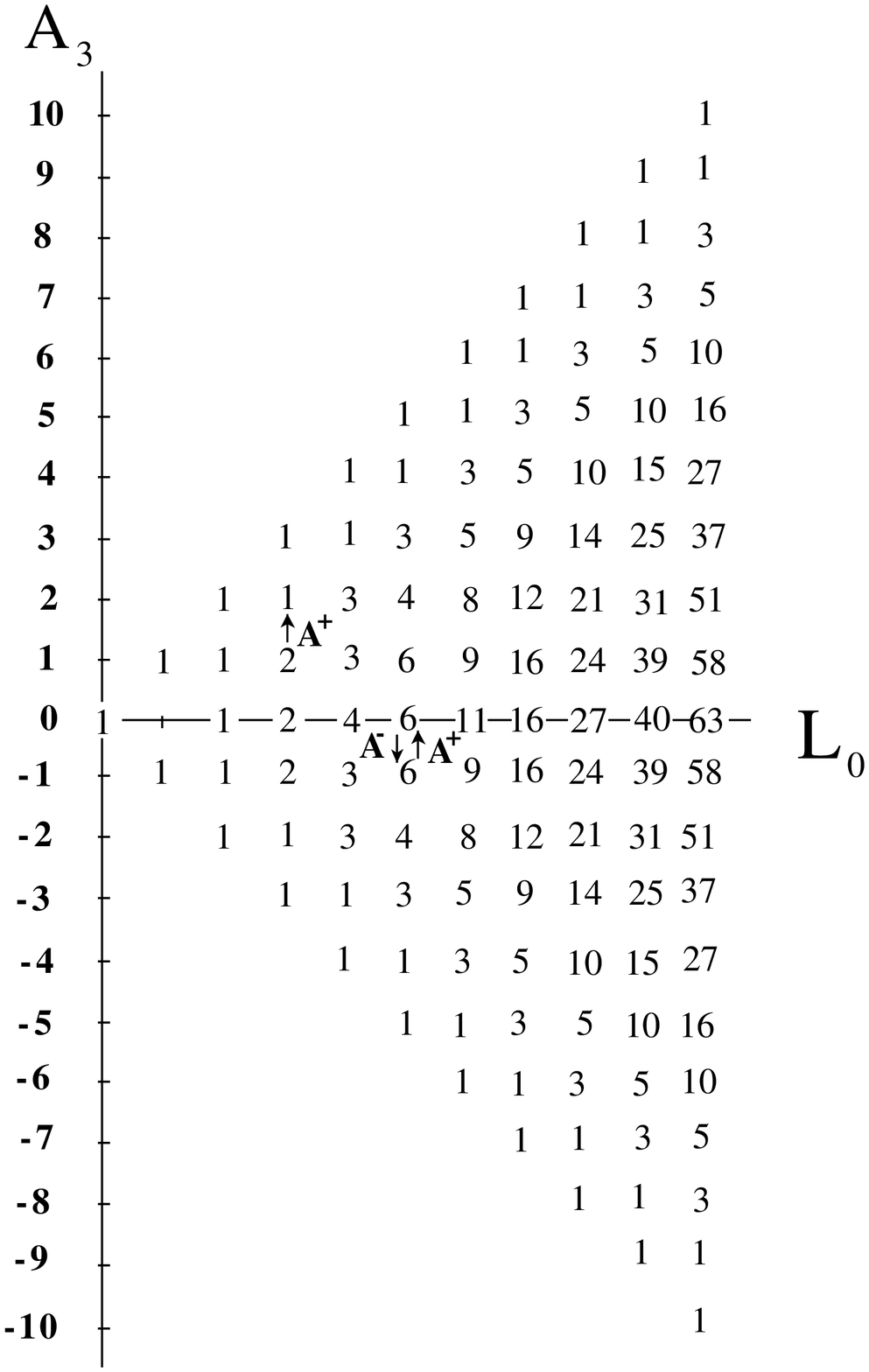}
\quad\quad \fignum Spectrum $(L_{0},A_{3})$.
\end{center}
\end{figure}
}

\noindent{\it Gauge 2.} Operators $A^{+}A^{-},A^{-}A^{+}$ act in subspaces
with fixed $(L_{0},A_{3}=0)$. Computing matrix elements
of these operators, and solving eigenvalue problem for
their symmetrical combination $C=(A^{+}A^{-}+A^{-}A^{+})/2$,
we obtain the spectrum $T_{i}=\sqrt{C_{i}}$, shown on \fref{f14}, left.
The eigenvectors of $C$ are presented in Table~3. 

\begin{figure}\label{f22}
\begin{center}
~\epsfysize=5cm\epsfxsize=5cm\epsffile{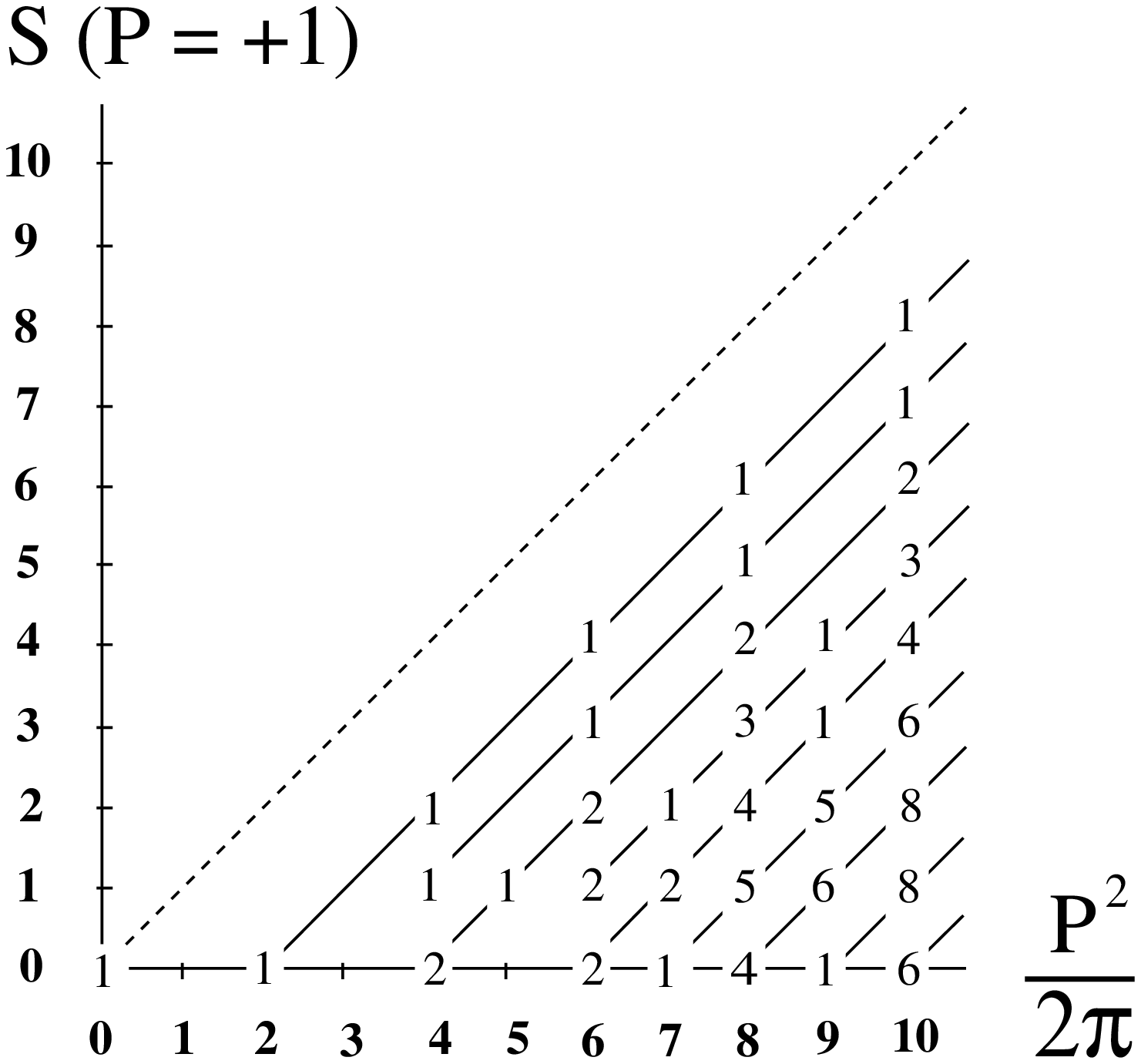}\quad\quad
~\epsfysize=5cm\epsfxsize=5cm\epsffile{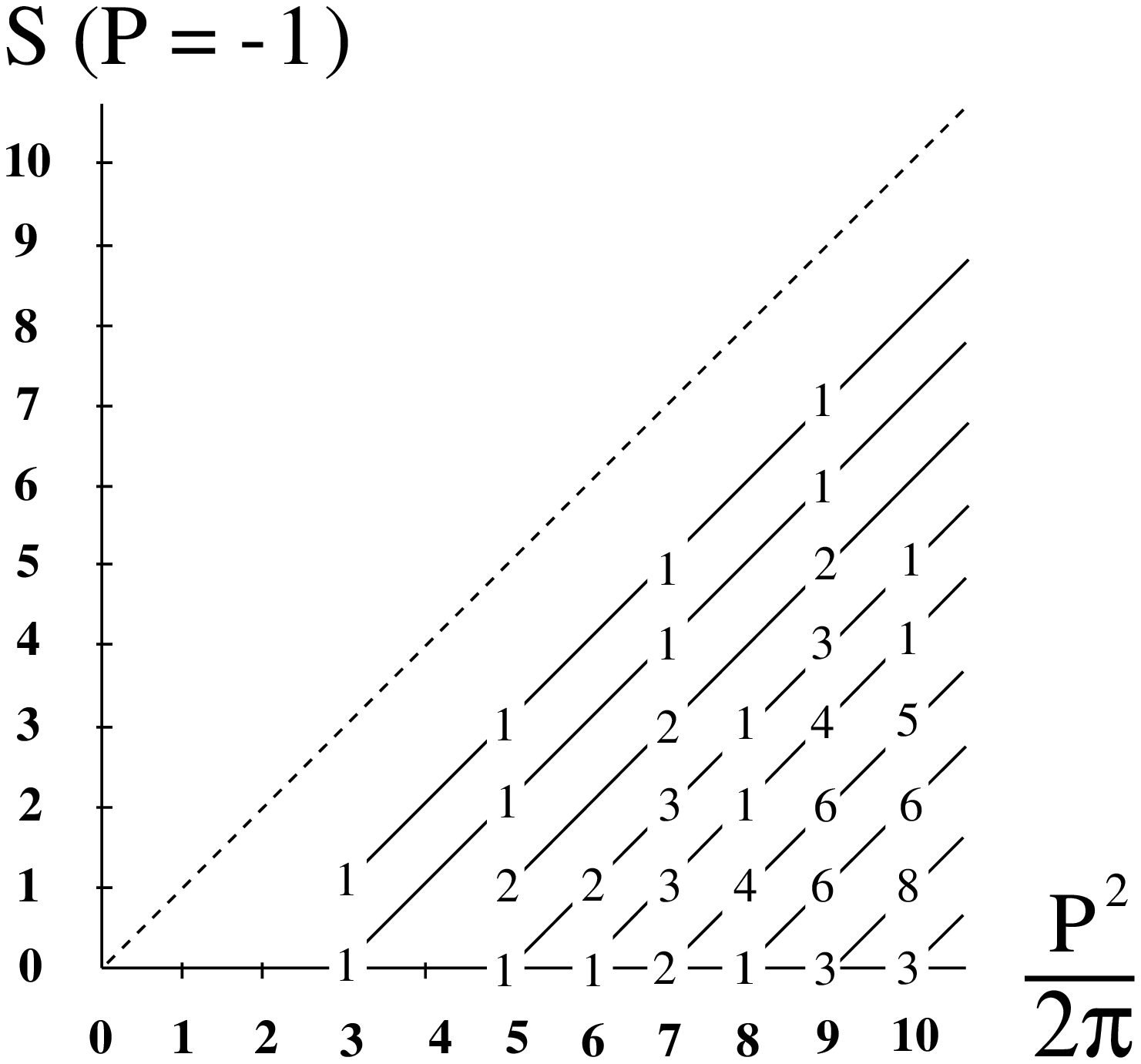}

\fignum Spin-mass spectra for the states with definite $P$-parity (gauge 2).
\end{center}
\end{figure}

$\Pi$-operation, represented by a replacement $n(k)\to n(-k)$ 
in state vector, acts in subspaces with fixed $(L_{0},A_{3}=0)$
and commutes with $C$. As a result, all eigenvectors of $C$
have definite $\Pi$-parity. Multiplying $\Pi$-parity by the factor 
$(-1)^{S},\ S=[T]$, obtain $P$-parity. Separate spin-mass spectra
for the states with definite $P$-parity are shown on \fref{f22}.

\vspace{2mm}

\def\ran{\rangle}

\begin{center}

{\scriptsize
$$\begin{array}{|c|l||c|l||c|l|}\hline
L_{0}S&|\{n(k)\}\ran&L_{0}S&|\{n(k)\}\ran&L_{0}S&|\{n(k)\}\ran\\\hline
00&|0\ran&40&|2(1)2(-1)\ran&51&|3(1)2(-1)\ran\\  
11&|1(1)\ran& &|1(-1)1(3)\ran& &|1(1)1(-1)1(3)\ran\\ 
20&|1(1)1(-1)\ran& &|1(1)1(-3)\ran&&|2(1)1(-3)\ran\\  
22&|2(1)\ran&42&|3(1)1(-1)\ran&&|1(5)\ran\\
31&|2(1)1(-1)\ran&&|1(1)1(3)\ran&53&|4(1)1(-1)\ran\\ 
&|1(3)\ran&44&|4(1)\ran&&|2(1)1(3)\ran\\ 
33&|3(1)\ran&&&55&|5(1)\ran\\\hline
\end{array}$$
}

{\bf Table 1:} basis of physical subspace for gauge 1,
axially symmetrical configurations.

{\scriptsize
$$\begin{array}{|c|l||c|l|}\hline
L_{0}S&|\{n(k)\}\ran&L_{0}S&|\{n(k)\}\ran\\\hline
00&|0\ran& 51&24|3(1)2(-1)\ran+9|1(-1)2(2)\ran-\\ 
11&|1(1)\ran&&96|1(1)1(2)1(-2)\ran+16|1(1)1(-1)1(3)\ran+\\
22&|2(1)\ran& &120|2(1)1(-3)\ran;\ 32|3(1)2(-1)\ran+7|1(-1)2(2)\ran+\\
31&9|2(1)1(-1)\ran+2|1(3)\ran&&32|1(1)1(2)1(-2)\ran+8|1(1)1(-1)1(3)\ran
+8|1(5)\ran\\  
33&|3(1)\ran&52&-12|2(1)1(-1)1(2)\ran+24|3(1)1(-2)\ran-\\
40&9|2(1)2(-1)\ran-18|1(2)1(-2)\ran+&&4|1(2)1(3)\ran+3|1(1)1(4)\ran\\ 
&5|1(-1)1(3)\ran+27|1(1)1(-3)\ran&53&2|4(1)1(-1)\ran+|1(1)2(2)\ran;\\
42&8|3(1)1(-1)\ran+3|2(2)\ran;&&9|4(1)1(-1)\ran+4|2(1)1(3)\ran\\ 
&3|3(1)1(-1)\ran+|1(1)1(3)\ran& 55&|5(1)\ran\\ 
44&|4(1)\ran&&\\\hline
\end{array}$$
}

{\bf Table 2:} basis of physical subspace for gauge 1,
Gupta-Bleuler's approach.

\vspace{3mm}
{\scriptsize
$$\begin{array}{|c|c|l|}\hline
L_{0}SP&T&|\{n(k)\}\ran\\\hline
0 0 +&0&|0\ran\\
2 0 +& 0.5 &|1(1)1(-1)\ran\\
3 0 -& 0.645 &0.707(|1(-1)1(2)\ran-|1(1)1(-2)\ran)\\
3 1 -& 1.48 &0.707(|1(-1)1(2)\ran+|1(1)1(-2)\ran)\\
4 0 +& 0.768 &0.630|2(1)2(-1)\ran-0.460|1(2)1(-2)\ran+\\
&&0.442(|1(-1)1(3)\ran+|1(1)1(-3)\ran)\\
4 0 +& 0.426 &0.633|2(1)2(-1)\ran+0.771|1(2)1(-2)\ran-\\
&&0.0496(|1(-1)1(3)\ran+|1(1)1(-3)\ran)\\
4 1 +& 1.63 &0.707(|1(-1)1(3)\ran-|1(1)1(-3)\ran)\\
4 2 +& 2.43 &-0.450|2(1)2(-1)\ran+0.440|1(2)1(-2)\ran+\\
&&0.549(|1(-1)1(3)\ran+|1(1)1(-3)\ran)\\
5 0 -& 0.64 &0.424(|1(1)2(-1)1(2)\ran-|2(1)1(-1)1(-2)\ran)+\\
&&0.558(|1(-2)1(3)\ran-|1(2)1(-3)\ran)+\\
&&0.0972(|1(-1)1(4)\ran-|1(1)1(-4)\ran)\\
5 1 +& 1.09 &-0.450(|1(1)2(-1)1(2)\ran-|2(1)1(-1)1(-2)\ran)+\\
&&0.406(|1(-2)1(3)\ran-|1(2)1(-3)\ran)-\\
&&0.364(|1(-1)1(4)\ran-|1(1)1(-4)\ran)\\
5 1 -& 1.41 &0.210(|1(1)2(-1)1(2)\ran+|2(1)1(-1)1(-2)\ran)+\\
&&0.633(|1(-2)1(3)\ran+|1(2)1(-3)\ran)-\\
&&0.235(|1(-1)1(4)\ran+|1(1)1(-4)\ran)\\
5 1 -& 1.61 &0.509(|1(1)2(-1)1(2)\ran+|2(1)1(-1)1(-2)\ran)+\\
&&0.0133(|1(-2)1(3)\ran+|1(2)1(-3)\ran)+\\
&&0.490(|1(-1)1(4)\ran+|1(1)1(-4)\ran)\\
5 2 -& 2.62 &0.343(|1(1)2(-1)1(2)\ran-|2(1)1(-1)1(-2)\ran)-\\
&&0.156(|1(-2)1(3)\ran-|1(2)1(-3)\ran)-\\
&&0.598(|1(-1)1(4)\ran-|1(1)1(-4)\ran)\\
5 3 -& 3.37 &0.443(|1(1)2(-1)1(2)\ran+|2(1)1(-1)1(-2)\ran)-\\
&&0.315(|1(-2)1(3)\ran+|1(2)1(-3)\ran)-\\
&&0.452(|1(-1)1(4)\ran+|1(1)1(-4)\ran)\\\hline
\end{array}$$
}

{\bf Table 3:} basis of physical subspace for gauge 2.
\end{center}

\vspace{3mm}\noindent
\underline{{\it Remarks.}}
In Tables~1,2 $|\{n(k)\}\ra>=\prod\limits_{k>0}
(a_{k}^{+})^{n(k)}(a_{-k})^{n(-k)}|0\ra>$. Norms of these states
can be calculated by the formula 
$N=\la<\{n(k)\}|\{n(k)\}\ra>=\prod\limits_{k>0}
k^{n(k)}n(k)!~k^{n(-k)}n(-k)!$
In Table~3 the states $|\{n(k)\}\ra>$ are {\it normalized}: 
$|\{n(k)\}\ra>=N^{-1/2}\prod\limits_{k>0}
(a_{k}^{+})^{n(k)}(a_{-k})^{n(-k)}|0\ra>$.
Longer versions of these tables, continued up to values
$L_{0}=10$, can be found in Internet under URL:
{\tt http://viswiz.gmd.de/\~{}nikitin/str/lcg.html}

\paragraph*{Appendix 4: Singularity on leading Regge trajectory.}\quad 

\vspace{1mm}\noindent
Let's consider the projection of supporting curve in CMF
and parametrize it by it's length: ${\bf Q}(L)$.
In this parametrization $dQ_{0}/dL=|d{\bf Q}/dL|=1$
(consequently, total length of the curve is equal to double mass of 
the string). Let's consider a point on the supporting curve, 
in which tangent vector $d{\bf Q}/dL$ is directed opposite to the gauge axis 
${\bf e}_{3}$, see \fref{f19}. In the vicinity of this point
the following expansion is valid:
${{d{\bf Q}}/{dL}}={\bf T}+{\bf N}(L-L^{*})+O((L-L^{*})^{2}),$
where ${\bf T}={{d{\bf Q}}/{dL}}=-{\bf e}_{3}$
is unit tangent vector to the curve in the point ${\bf Q}(L^{*})$, and 
${\bf N}={{d^{2}{\bf Q}}/{dL^{2}}}\perp {\bf e}_{3}$
is a major normal to the curve in this point. Thus, the expansion of
$(dQ_{0}/dL+dQ_{3}/dL)$ starts from $(L-L^{*})^{2}$.

On the other hand, in light cone gauge (\ref{conf1})
$dQ_{0}/d\s+dQ_{3}/d\s={{dL}/{d\sigma}}\left({{dQ_{0}}/{dL}}+
{{dQ_{3}}/{dL}}\right)={\sqrt{P^{2}}/{\pi}}=Const,$
therefore ${{dL}/{d\sigma}}\sim(L-L^{*})^{-2}\ \Rightarrow\
(L-L^{*})\sim(\s-\s^{*})^{1/3}$. Then $dQ_{0}/d\s=dL/d\s
\sim(\s-\s^{*})^{-2/3},\ dQ_{3}/d\s={\sqrt{P^{2}}/{\pi}}-dQ_{0}/d\s
\sim(\s-\s^{*})^{-2/3}$, and $(dQ_{1}/d\s)^{2}+(dQ_{2}/d\s)^{2}
=2\sqrt{P^{2}}/{\pi}\cdot dQ_{0}/d\s-P^{2}/\pi^{2}\sim(\s-\s^{*})^{-2/3}$.

\vspace{-0.5mm}
\hfill\parbox[t]{6.3cm}{\begin{center}\begin{figure}\label{f19}
~\epsfysize=4cm\epsfxsize=5cm\epsffile{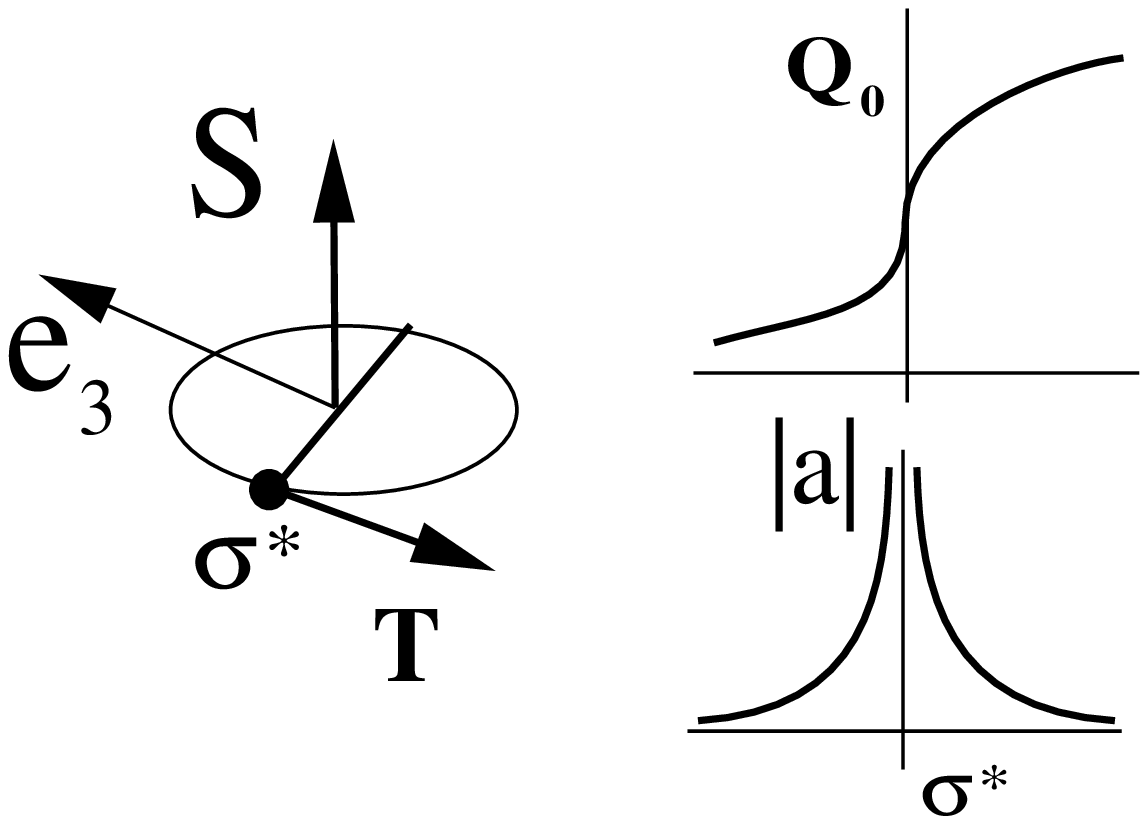}

\fignum Singularity of light cone gauge.
\end{figure}\end{center}
}\quad
\parbox[t]{5.2cm}{
Thus, the light cone gauge parameterizes the supporting curve 
irregularly in the vicinity of point $Q_{\mu}(\sigma^{*})$. At $\s\to\s^{*}$
the components of tangent vector $Q_{\mu}'(\s)$ tend to infinity:
$Q_{0,3}'(\s)$ as $(\s-\s^{*})^{-2/3}$, and $Q_{1,2}'(\s)$ as 
$(\s-\s^{*})^{-1/3}$. This type of singularity is integrable,
it does not create any problems for classical theory.
Particularly, masses of such configurations are finite.
However, such configurations require an infinite number 
of excited modes in Fourier expansion of $a(\s)=Q_{1}'(\s)-iQ_{2}'(\s)$.
This makes problematic a consideration of such configurations
in quantum theory, where the states with finite mass necessarily have
finite number of filled modes. 
}

\vspace{2.5mm}
For string theory in $d=4$ the described singularity
is not crucial, because the alignment of ${\bf Q}'(\sigma^{*})
\uparrow\downarrow{\bf e}_{3}$ can be removed 
by a small deformation of the curve, so
the configurations with such singularity are rare in the whole phase space.
(This singularity is a real obstacle for $d=3$. In 2-dimensional CMF
such alignment cannot be removed by a small deformation, and
most of string configurations become infinite-modal in light cone gauge.)

In the case, if the gauge axis is selected perpendicular to the spin
(gauge 2), the described singularity appears on the leading Regge trajectory.
Leading trajectory
corresponds to circular supporting curves (straight strings \cite{slstring}),
and for any direction of the gauge axis, perpendicular to the spin,
a point on the circle exists with tangent opposite to the gauge axis.
This singularity is a possible reason for
disappearance of the leading trajectory in quantization.

\baselineskip=\normalbaselineskip\normalsize

\nonumsection{References}

\end{document}